\documentclass[a4paper,11pt]{article}
  \pdfoutput=1 

\usepackage{jheppub} 

\usepackage[T1]{fontenc} \usepackage{amssymb}
\usepackage[colorinlistoftodos,prependcaption,textsize=tiny]{todonotes}

\usepackage[utf8]{inputenc}
\DeclareUnicodeCharacter{27E8}{\ensuremath{\langle}}
\DeclareUnicodeCharacter{27E9}{\ensuremath{\rangle}}
\usepackage{amsmath, amsthm, mathtools}
\usepackage{hyperref}
\usepackage{multicol,multirow}
\usepackage[sort&compress,numbers]{natbib}
\usepackage{subfigure}
\usepackage{booktabs}

\newcommand\fverb{\setbox\fverbbox=\hbox\bgroup\verb}
\newcommand\fverbdo{\egroup\medskip\noindent \fbox{\unhbox\fverbbox}\ }
\newcommand\fverbit{\egroup\item[\fbox{\unhbox\fverbbox}]}
\newbox\fverbbox

\newcommand{\dRleplep}{\ensuremath{\Delta R_{\ell\ell}}}
\newcommand{\mtw}{\ensuremath{m_{T,W}}}
\newcommand{\ptmissrel}{\ensuremath{p_{T}^{\text{miss,rel}}}}
\newcommand{\etal}{\ensuremath{\eta_{\ell}}}

\newcommand{\ptltwo}{\ensuremath{p_{T,\ell_2}}}
\newcommand{\ptlone}{\ensuremath{p_{T,\ell_1}}}
\newcommand{\dRlepgamma}{\ensuremath{\Delta R_{\ell\gamma}}}
\newcommand{\dRgammajet}{\ensuremath{\Delta R_{\gamma j}}}
\newcommand{\dRlepjet}{\ensuremath{\Delta R_{\ell j}}}
\newcommand{\ptgamma}{\ensuremath{p_{T,\gamma}}}
\newcommand{\ptjet}{\ensuremath{p_{T,j}}}
\newcommand{\ptmiss}{\ensuremath{p_{T}^{\text{miss}}}}
\newcommand{\etagamma}{\ensuremath{|\eta_{\gamma}|}}
\newcommand{\ptgammatwo}{\ensuremath{p_{T,\gamma_2}}}
\newcommand{\ptgammaone}{\ensuremath{p_{T,\gamma_1}}}
\newcommand{\ptlep}{\ensuremath{p_{T,{\ell}}}}
\newcommand{\etalep}{\ensuremath{|\eta_{\ell}|}}
\newcommand{\mll}{\ensuremath{m_{\ell^-\ell^+}}}
\def\psing{p_{{\rm singlet}}}
\newcommand{\rd}{{\rm d}}
\def\tauzero{{\cal T}_0}

\def\shat{\hat{s}}
\def\that{\hat{t}}
\def\uhat{\hat{u}}

\def\Bcal{\cal B}
\def\tauzero{{\cal T}_0}

\def\beq{\begin{equation}}
\def\eeq{\end{equation}}

\def\I33m{\mathrm{I}_3^{3{\mathrm m}}}
\def\nn{\nonumber}

\def\be{\begin{equation}}
\def\ee{\end{equation}}
\def\beqn{\begin{eqnarray}}
\def\eeqn{\end{eqnarray}}
\def\bea{\begin{eqnarray}}
\def\eea{\end{eqnarray}}

\def\spa#1.#2{\left\langle#1#2\right\rangle}
\def\spb#1.#2{\left[#1#2\right]}
\def\lor#1.#2{\left(#1#2\right)}
\def\sand#1.#2.#3{\left\langle\smash{#1}{\vphantom1}^{-}\right|{#2}\left|\smash{#3}{\vphantom1}^{-}\right\rangle}
\def\sandp#1.#2.#3{\left\langle\smash{#1}{\vphantom1}^{-}\right|{#2}\left|\smash{#3}{\vphantom1}^{+}\right\rangle}
\def\sandpp#1.#2.#3{\left\langle\smash{#1}{\vphantom1}^{+}\right|{#2}\left|\smash{#3}{\vphantom1}^{+}\right\rangle}
\def\sandpm#1.#2.#3{\left\langle\smash{#1}{\vphantom1}^{+}\right|{#2}\left|\smash{#3}{\vphantom1}^{-}\right\rangle}
\def\sandmp#1.#2.#3{\left\langle\smash{#1}{\vphantom1}^{-}\right|{#2}\left|\smash{#3}{\vphantom1}^{+}\right\rangle}
\def\spab#1.#2.#3{\langle#1|#2|#3]}
\def\spba#1.#2.#3{[#1|#2|#3\rangle}
\def\spaaold#1.#2.#3{\langle#1|#2|#3\rangle}
\def\spbbold#1.#2.#3{[#1|#2|#3]}
\def\spaa#1.#2.#3.#4{\left\langle#1|#2|#3|#4\right\rangle}
\def\spbb#1.#2.#3.#4{\left[#1|#2|#3|#4\right]}
\def\spaxa#1.#2.#3.#4{\langle#1|#2|#3|#4\rangle}
\def\spbxb#1.#2.#3.#4{[#1|#2|#3|#4]}

\def\prp34{\mathcal{P}_{34}}
\def\prpp345{\mathcal{P}_{345}}

\usepackage{slashed}
\allowdisplaybreaks

\preprint{\begin{minipage}[t]{8cm}\begin{flushright}FERMILAB-PUB-22-075-T,\\ IPPP/22/05\end{flushright}\end{minipage}}
\title{Non-local slicing approaches for NNLO QCD in MCFM}

\author[a]{John M. Campbell,}
\emailAdd{johnmc@fnal.gov}

\author[b]{R. Keith Ellis,}
\emailAdd{keith.ellis@durham.ac.uk}

\author[c]{Satyajit Seth}
\emailAdd{seth@prl.res.in}

\affiliation[a]{Fermilab, PO Box 500, Batavia IL 60510-5011, USA}
\affiliation[b]{Institute for Particle Physics Phenomenology, Durham University, Durham, DH1 3LE, UK}
\affiliation[c]{Physical Research Laboratory, Navrangpura, Ahmedabad - 380009, India}
\date{\today}

\abstract{
We present the implementation of several processes at
Next-to-Next-to Leading Order (NNLO) accuracy in QCD
in the parton-level Monte Carlo program MCFM.
The processes treated are $pp\to H$, $W^\pm$, $Z$, $W^\pm H$, $ZH$,
$W^\pm\gamma$, $Z\gamma$ and $\gamma\gamma$ and, for the first time
in the code, $W^+W^-$, $W^\pm Z$ and $ZZ$.
Decays of the unstable bosons are fully included,
resulting in a flexible fully differential Monte Carlo code.
The NNLO corrections have been calculated using two non-local
slicing approaches, isolating the doubly unresolved region
by cutting on the zero-jettiness, $\tauzero$, or on $q_T$, the transverse momentum 
of the colour singlet final-state particles.
We find that for most, but not all processes the $q_T$ slicing method
leads to smaller power corrections for equal computational burden.}


\begin{document} 
\setcounter{tocdepth}{2}
\maketitle

\section{Introduction}

The current and future success of the LHC depends crucially on the
precision supplied by theoretical calculations.  Reducing the
theoretical error has special importance in the context of Higgs
Physics\cite{Cepeda:2019klc} where for many channels it is projected
to remain the largest error at the conclusion of the LHC program.

An important role is played by processes involving the bosons, $W,Z,\gamma$ and $H$.
The importance of detailed studies of the Higgs boson goes without saying.
Single vector boson production can be used as a luminosity monitor and to probe
parton distribution functions.
Electroweak production of vector boson pairs is a stringent test of the
standard model. In addition the $\gamma\gamma$, $ZZ$, $WW$ and $Z\gamma$ processes have renewed significance
because they constitute backgrounds to Higgs boson decay processes.

We are preparing a release of MCFM~\cite{Campbell:1999ah,Campbell:2011bn,Campbell:2015qma,Campbell:2019dru}
which will allow calculation of the NNLO QCD results for a
large number of colour singlet production processes with both
zero-jettiness~\cite{Stewart:2010tn,Boughezal:2015dva,Gaunt:2015pea} and
$q_T$-slicing~\cite{Catani:2007vq}.\footnote{We aim to release this version in March 2022.} 
The code has also been used
for processes with non-colour singlet final states, such
as $W$~+ jet~\cite{Boughezal:2015dva}, $Z$~+jet~\cite{Boughezal:2015ded},
$\gamma$~+jet~\cite{Campbell:2016lzl}, and Higgs boson + jet~\cite{Campbell:2019gmd,Mondini:2021nck},
although these have not yet
been made available in the public version.
These latter processes are treatable 
because of the SCET factorization theorems for the cross sections for small 1-jettiness.

Given the importance of precision for the LHC, 
there has been an intense community effort to produce results
at NNLO QCD and in some cases N$^3$LO.
A review of the 2020 status of precision QCD with a special focus on the Higgs boson is given in Ref.~\cite{Heinrich:2020ybq}.
In Table~\ref{NNLOcites} we present references for the processes that have been calculated in NNLO QCD.
\begin{table}[b]
\begin{center}
\begin{tabular}{|l|l|l|l|}
\hline
Process & MCFM & Process & MCFM \\
\hline
$H+ 0$~jet~\cite{Anastasiou:2004xq,Anastasiou:2005qj,Catani:2007vq,Gaunt:2015pea,Mistlberger:2018etf,Cieri:2018oms,Chen:2021isd}  & \checkmark~\cite{Boughezal:2016wmq}
&$W^\pm+ 0$~jet~\cite{Melnikov:2006di,Catani:2009sm,Grazzini:2017mhc} & \checkmark~\cite{Boughezal:2016wmq} \\
$Z/\gamma^* + 0$~jet~\cite{Melnikov:2006kv,Catani:2009sm,Gaunt:2015pea,Grazzini:2017mhc} & \checkmark~\cite{Boughezal:2016wmq}
& $ZH$~\cite{Ferrera:2014lca} & \checkmark~\cite{Campbell:2016jau} \\
$W^\pm\gamma$~\cite{Grazzini:2015nwa,Grazzini:2017mhc,Cridge:2021hfr} & \checkmark~\cite{Campbell:2021mlr}
& $Z\gamma$~\cite{Grazzini:2017mhc,Campbell:2017aul}&  \checkmark~\cite{Campbell:2017aul} \\
$\gamma \gamma$~\cite{Catani:2011qz,Grazzini:2017mhc,Catani:2018krb,Alioli:2020qrd}&  \checkmark~\cite{Campbell:2016yrh} 
&single top~\cite{Brucherseifer:2014ama}  & \checkmark~\cite{Campbell:2020fhf} \\
$W^\pm H$~\cite{Ferrera:2013yga,Caola:2017xuq} & \checkmark~\cite{Campbell:2016jau} & $WZ$~\cite{Grazzini:2016swo,Grazzini:2017ckn}              &  \checkmark\\
$ZZ$~\cite{Cascioli:2014yka,Grazzini:2015hta,Caola:2015psa,Heinrich:2017bvg,Grazzini:2017mhc,Kallweit:2018nyv,Grazzini:2018owa} & \checkmark
&$W^+W^-$~\cite{Gehrmann:2014fva,Caola:2015rqy,Grazzini:2016ctr,Grazzini:2017mhc,Grazzini:2020stb}  & \checkmark \\
$W^{\pm} + 1$~jet~\cite{Boughezal:2016dtm,Gehrmann-DeRidder:2017mvr}& \cite{Boughezal:2015dva}
& $Z+ 1$~jet~\cite{Gehrmann-DeRidder:2015wbt,Gehrmann-DeRidder:2016cdi} &  \cite{Boughezal:2015ded}\\
$\gamma+ 1$~jet~\cite{Chen:2019zmr} & \cite{Campbell:2016lzl}
& $H+ 1$~jet~\cite{Boughezal:2013uia,Chen:2014gva,Boughezal:2015dra,Boughezal:2015aha,Caola:2015wna,Chen:2016vqn} & \cite{Campbell:2019gmd}\\
$b\bar b \to H+$jet& \cite{Mondini:2021nck} & &\\
$t \bar{t}$~\cite{Czakon:2016ckf,Abelof:2015lna,Catani:2019hip,Catani:2019iny,Czakon:2020qbd,Catani:2020tko}  &
& $Z+b$~\cite{Gauld:2020deh}& \\
$W^{\pm}H$+jet~\cite{Gauld:2020ced}& & $ZH$+jet~\cite{Gauld:2021ule}&  \\
Higgs WBF~\cite{Cacciari:2015jma,Buckley:2021gfw}  & & $H \to b \bar{b}$~\cite{Anastasiou:2011qx,DelDuca:2015zqa,Mondini:2019gid}& \\
top decay~\cite{Gao:2012ja,Brucherseifer:2013iv,Campbell:2020fhf}  & & dijets~\cite{Currie:2014upa,Currie:2016bfm,Gehrmann-DeRidder:2019ibf}  & \\
$\gamma \gamma$+jet~\cite{Chawdhry:2021hkp}&  & $W^{\pm}c$~\cite{Czakon:2020coa}&\\
$b\bar{b}$~\cite{Catani:2020kkl}& & $\gamma \gamma \gamma$~\cite{Chawdhry:2019bji,Kallweit:2020gcp} & \\
HH~\cite{Grazzini:2018bsd}& &HHH~\cite{deFlorian:2019app}&\\
\hline
\end{tabular}
\end{center}
\caption{Publications on processes evaluated differentially at NNLO, (and in some cases beyond NNLO). The tick mark indicates
  that the process is available in the public MCFM version. Processes with a reference but no tickmark are not yet in the public MCFM code.
Processes with a tickmark but no reference have been introduced into the public code at this time.}
\label{NNLOcites} 
\end{table}
 It is important to note that when targeting the precision achievable at NNLO, electroweak corrections can also become important,
especially at large $p_T$. A discussion of these effects is beyond the scope of this paper.

In Ref.~\cite{Boughezal:2016wmq} MCFM results for $pp\to H$, $pp\to Z$,
$pp\to W$, $pp\to ZH$, $pp\to WH$ and $pp\to \gamma \gamma$ have been
presented.  Results for colour singlet production processes,
especially vector boson pairs have also been presented by the MATRIX
collaboration~\cite{Grazzini:2017mhc,Grazzini:2019jkl}.  Therefore,
although results for the colour singlet cross sections presented in
this paper are known, in view of the complicated nature of these
calculations it is re-assuring to have an independent check. 
The results of Ref.~\cite{Grazzini:2017mhc,Grazzini:2019jkl} calculate one-loop
virtual corrections using Openloops~2~\cite{Buccioni:2019sur}, whereas
in MCFM the one-loop virtual corrections are calculated analytically,
with consequent benefits for the stability and speed of this portion
of the code.  Note however that the MATRIX and MCFM calculations for
vector boson pairs can not be considered totally independent, relying
as they do on the same two-loop
amplitudes~\cite{Gehrmann:2011ab,Gehrmann:2015ora}.  We also compare
with an calculation of $ZZ$ production~\cite{Heinrich:2017bvg}, which is independent
(except for the same caveat about two-loop matrix elements).

Recently there has been a detailed re-examination  
of fiducial cross sections for two-body
decay processes at colliders, demonstrating that certain commonly used cuts
are sensitive to low momentum scales~\cite{Alekhin:2021xcu,Salam:2021tbm}.\footnote{
Methods to remove the dominant (linear) sensitivity in the low-momentum region
in a fully differential way, for the $q_T$-slicing method,
have been developed in Refs.~\cite{Buonocore:2021tke,Camarda:2021jsw}.}
Where possible we shall limit our discussion in this paper to total inclusive cross sections,
leaving detailed predictions with well-motivated cuts to a subsequent paper.  

A successive improvement of our results could come from widespread inclusion
of resummation effects along the lines of Refs.~\cite{Becher:2020ugp,Neumann:2021zkb}.
Resummation effects in vector boson pair production have previously been considered in Refs.~\cite{Grazzini:2015wpa,Kallweit:2020gva,Wiesemann:2020gbm}.
Resummation is also part of the program in the GENEVA collaboration. Ref.~\cite{Alioli:2021ggd} provides
a recent article, where references to earlier work of the GENEVA collaboration can be found.

\section{Non-local slicing methods}
\label{sec:methods}
In this section we review the calculation of the NNLO cross sections
which have colour singlet final states at the Born level.
The necessary requisites for the methods are,
\begin{itemize}
\item An analytic understanding of the behaviour of the Born process
  accompanied by soft and collinear radiation through to the requisite
  order, i.e.~for NNLO through to order $\alpha_s^2$.
\item A NLO calculation of the process at hand with one additional
  parton.
\item The two-loop virtual corrections to the process at hand,
  necessary to calculate the hard function at order $\alpha_s^2$
\end{itemize}

Let $r$ be a zero jet resolution variable which divides the phase space in two,
\beq
\sigma(X)=\sigma(X,r^{cut})+\int_{r^{cut}}\; dr^\prime \; \frac{d \sigma(X)}{d r^\prime},
\eeq
where $X$ represents other kinematics on the phase space.
In the following subsections we shall take the resolution variable, $r^{cut}$
to be either the 0-jettiness, $\tauzero$ or the transverse momentum, $q_T$ of the final state.
\beq
\sigma(X,r^{cut})=\int^{r^{cut}} \; dr^\prime \; \frac{d \sigma(X)}{d r^\prime}.
\eeq
The cross section is given by introducing $\sigma^{sub}$, the analytic form for the  
cross section, known for small values of the resolution parameter $r^{cut}$ from
factorization theorems.
\beqn
\sigma&=&
\sigma^{sub}(r^{cut})+\int^{r^{cut}} \; d r^\prime \; \frac{d \sigma(X)}{d r^\prime}
  +[\sigma(r^{cut})-\sigma^{sub}(r^{cut})] \nonumber \\
&\equiv& \sigma^{sub}(r^{cut})+\int^{r^{cut}} \; d r^\prime \; \frac{d \sigma(X)}{d r^\prime}
  +[\Delta \sigma(r^{cut})].
\eeqn
Since $r$ is a zero jet resolution variable, $\Delta \sigma(r^{cut})$ will tend to zero as
$r^{cut} \to 0$. 

\subsection{Non-local jettiness slicing}
\label{sec:SCET}

The $0$-jettiness slicing method is based on the corresponding event
shape introduced in Ref.~\cite{Stewart:2010tn}.
Writing $q^\mu, Q$ and $Y$ for the four-momentum, mass and rapidity
of the colour singlet system in its centre of mass, the incoming parton momenta are
\beq
p_i = x_i E_{cm} \frac{n}{2},\;\; p_j = x_j E_{cm} \frac{\bar{n}}{2}, 
\eeq
where $n=(1,+\vec{z}),\bar{n}=(1,-\vec{z})$. The zero-jettiness in the
colour singlet centre of mass is then defined by,
\beq \label{tau0CM}
\tauzero = \sum_k \min \{ e^{+Y} p_k^+, e^{-Y} p_k^- \},
\eeq
where the sum over $k$ runs over all final state partons and $p_k^- = p.n$,
$p_k^+ = p.\bar{n}$.
The all-orders resummed form of the cross section in the region of small $\tauzero$,
obtained by application of soft-collinear 
effective theory (SCET)~\cite{Bauer:2000ew,Bauer:2000yr,Bauer:2001ct,Bauer:2001yt,Bauer:2002nz},
is then given by, 
\begin{equation}
\label{eq:SCETfac1}
\frac{\rd \sigma}{\rd \tauzero} =\sum_{ij} \int \rd x_i \rd x_j \int \rd
\Phi_B(p_i,p_j;\psing)\, H_{ij}(\Phi_B,\mu)\,\frac{\rd\Delta_{ij}}{\rd\tauzero}
+\ldots\ ,
\end{equation}
where the indices $i,j$ run over all initial state partons involved in the
scattering.  
$\Phi_B$ represents the Born-level color singlet phase space $p_i p_j\rightarrow\psing$
and $H_{ij}$ the hard function. 
The soft/collinear function $\Delta_{ij}$ is,
\begin{eqnarray}
\label{eq:SCETfac2}
\frac{\rd\Delta_{ij}}{\rd\tauzero}&=&B_{i/H_1}\otimes B_{j/H_2}\otimes S_{ij}\nonumber\\
&\equiv& \int \rd t_{B_i} \rd t_{B_j}  \rd t_S \, \delta\left(
 \tauzero-t_{B_i}-t_{B_j}-t_S\right) 
 \, B_{i/H_1}(t_{B_i},x_i) \, B_{j/H_2}(t_{B_j},x_j)
         \,S_{ij}(t_S)\,.
\end{eqnarray}
The hard function encodes both the leading order matrix elements and
perturbative virtual corrections as described later in section~\ref{sec:hard}.
The beam function $B_{i/H}$ describes initial-state collinear radiation from hadron $H$
and can be written as a
convolution of perturbative matching coefficients and the usual parton
density functions, $f_{i/H}$. It has been computed up to two loops in Refs.~\cite{Gaunt:2014xga,Gaunt:2014cfa}.
The effects of soft radiation are collected in the soft function $S$, which has been calculated
for zero-jettiness up to two-loop order in Refs.~\cite{Kelley:2011ng,Monni:2011gb}.

In the color singlet centre of mass frame the 
power corrections to the factorization in Eq.~(\ref{eq:SCETfac1}) are known to
be reduced~\cite{Moult:2016fqy}.   Power corrections to the simplest $2 \to 1$ processes
are known~\cite{Boughezal:2016zws,Boughezal:2018mvf,Moult:2016fqy,Moult:2017jsg,Ebert:2018lzn}
but, since they are not known universally and we also wish to compare with the $q_T$ approach,
we do not include them in this study.

\subsection{Non-local $q_T$ slicing}
In this section we briefly describe the calculation using the transverse momentum as
a resolution parameter. Although the formalism we describe is not the formalism in which $q_T$ slicing was originally
implemented~\cite{Collins:1981uk, Collins:1984kg, Catani:2000vq, Ji:2004wu, Ji:2004xq, Bozzi:2005wk}
it is simplest to implement $q_T$ slicing using the factorized
form of the low $q_T$ cross section derived using SCET.  Schematically, the differential cross section
takes the form,
\begin{align}
  \label{eq:fac1}
  \frac{d^2\sigma}{dQ dq_T} \sim \tilde{\Bcal}_{i/H_1}(x_1,k_{1T},\mu;\xi_1) \otimes \tilde{\Bcal}_{j/H_2}(x_2,k_{2T},\mu;\xi_2)
  \otimes \tilde{\mathcal{S}}_{ij}(q_{T},\mu;\xi_1,\xi_2) \otimes H_{ij}(z,Q,\mu) \,,
\end{align}
where the symbol $\otimes$ denotes a convolution.  Note that the soft function $\tilde{\mathcal{S}}$ and the
naive transverse PDFs $\tilde{\Bcal}$ depend on unphysical parameters, $\xi_1$ and $\xi_2$. However,
in physical cross sections the dependence on these parameters appears in such a way that only the physical scale $Q$ remains.

The $\tilde{\Bcal}$ and $\tilde{\mathcal{S}}$ functions still depend on both $Q$ and $q_T$, two disparate scales.
As such, Eq.~(\ref{eq:fac1}) does not represent a true factorization and thus additional work must be performed to
isolate the dependence on the scale $Q$.
We follow the SCET re-factorization approach of \cite{Becher:2010tm, Becher:2012yn}.
Thus for the simplest Drell-Yan process we have,
\begin{align}
\frac{d^3\sigma}{dQ^2dq_T^2 d y}
=\;& \frac{\alpha^2}{3N_c Q^2 s}
 \sum_{i,j}
\sum_{q}
 e_q^2
\left[
C_{q\bar{q}\leftarrow ij}
( z_1,z_2,q_T^2,Q^2,\mu) + ( q\leftrightarrow \bar{q} ) \right]
\nonumber\\
&\hspace{3.1cm}\otimes f_{i/H_1}( z_1,\mu)\otimes f_{j/H_2}( z_2,\mu)
\,,
\label{eq:dsigma_DY_full}
\end{align}
which is correct up to power corrections in $q_T^2/Q^2$ and $x_T^2 \Lambda_\text{QCD}^2$, where
$x_T^2 = -x_\perp^2$ and $x_\perp$ is the Fourier conjugate variable to $q_T$.
The perturbative function $C_{q\bar{q}\leftarrow ij}$ is given in terms
of the Wilson coefficient $C_V$ as,
\begin{align}
C_{q\bar{q}\leftarrow ij}
( z_1,z_2,q_T^2,Q^2,\mu)
=\,&
\left| C_V( Q^2,\mu)\right|^2 \!
\int\!\!  d^2 \! x_\perp\, e^{-i q_\perp\!\cdot x_\perp}
\left(\frac{x_T^2q^2}{4e^{-2\gamma_E}} \right)^{\!\!\!-F_{q\bar{q}}(x_T^2,\mu)}
\nonumber\\
& \times
I_{q/ i}( z_1,x_T^2,\mu) I_{\bar{q}/ j}( z_2,x_T^2,\mu) \,.
\label{eq:Cqqb_DY}
\end{align}
This is the form in which we have implemented the factorization
formula, taking the explicit form for the functions $I_{q/ i}$ and $F$
from Ref.~\cite{Gehrmann:2014yya}.  Since all the formula are
explicitly given in a machine readable format this gives the simplest
implementation method.  An alternative approach would be to use
Ref.~\cite{Billis:2019vxg} where a number of useful results are
collected. Ref.~\cite{Billis:2019vxg} has the ambition to be more
complete, since it also gives results which are relevant at N$^3$LO,
but for our purposes, i.e. NNLO, it is less useful since for some of the needed
components it refers to other papers. We have checked that a full NNLO
implementation based on Ref.~\cite{Billis:2019vxg} and papers
referenced therein gives the same result as in
Ref.~\cite{Gehrmann:2014yya}.

The $q_T$ spectrum for color singlet production including the complete
$q_T^2/Q^2$ power corrections at $O(\alpha_s)$, has been presented in
ref.~\cite{Ebert:2018gsn}.  Power corrections for the NLO inclusive
cross section of Drell-Yan processes have been calculated up to fourth
order in a transverse-momentum cut in Ref.~\cite{Cieri:2019tfv}, and
up to second order for the NNLO $qg$-initiated channel in
Ref.~\cite{Oleari:2020wvt}.Therefore a full suite of power corrections
at NNLO is not available, even for the simplest processes.

\subsection{Hard functions}
\label{sec:hard}
The hard functions are related to finite parts of the virtual one- and
two-loop corrections to the Born process.  For the case of the
Drell-Yan type processes, ($W,Z$ and $\gamma^*$) the two-loop
corrections are given in Refs.~\cite{Matsuura:1987wt,Gehrmann:2005pd}.
For the case of Higgs production in the large $m_t$ limit the two-loop corrections are given in
Refs.~\cite{Dawson:1990zj,Djouadi:1991tka,Gehrmann:2005pd,Ahrens:2009cxz}.
For the $V\gamma$ ($V=Z, W^\pm$) processes the finite remainders of
the one-loop and two-loop form factors are given in
\cite{Gehrmann:2011ab}, while the remainders for the $\gamma\gamma$
process are specified in Ref.~\cite{Anastasiou:2002zn}.  For the
diboson ($W^+W^-,W^\pm Z,ZZ)$ processes, with the vector bosons
decaying leptonically, the matrix elements up to two loops have been
computed in Ref.~\cite{Gehrmann:2015ora}.  Details of the conversion
of the two-loop matrix elements in Ref.~\cite{Gehrmann:2015ora}
to the hard functions are presented
in Appendix~\ref{hardappendix}.  We employ
HandyG~\cite{Naterop:2019xaf} for the numerical evaluation of multiple
polylogarithms that appear in the expression of two-loop finite
remainders.
 
\subsection{Above cut contributions }
A necessary ingredient for the vector boson pair calculations is the NLO calculation of
the desired parton process but with one additional parton in the final state.
In preparation for this paper we have implemented and improved the treatment
of the $VV+$~jet process at one-loop, for the cases of $VV=W^+W^-,W^\pm Z,ZZ$.
As with all NLO
processes in MCFM these are included using analytic formula. Analytic results for the one-loop
calculation of the $W^+W^-$+3 parton process have been given in Ref.~\cite{Campbell:2015hya}.
After simple modifications these results can also be applied to the $WZ$ and $ZZ$ processes.
Although the results of Ref.~\cite{Campbell:2015hya} were in analytic form,
considerable effort has been devoted to simplifying these results~\cite{Campbell:2022qpq}.
Analytic results for the one-loop calculation of the $W^+\gamma$+3 parton process,
applicable also to the $Z\gamma$ process, have been given in Ref.~\cite{Campbell:2021mlr}.

We have compared our one-loop results with OpenLoops~2~\cite{Buccioni:2019sur}
and Recola2~\cite{Actis:2016mpe,Denner:2017wsf},
both as a confirmation of the results and to establish timings. The comparison is performed
via an extension of the C++ interface to MCFM~\cite{Campbell:2021vlt},
computing the interference of Born and 1-loop amplitudes using the same set of 1000 representative
phase-space points with the default setup for both OpenLoops~2 (version 2.1.2) and Recola2 (version 2.2.3).  Our
calculations agree perfectly with those of these libraries\footnote{
For the $Z\gamma+$jet and $WW+$jet processes agreement is established only in the
limit $m_t \to \infty$ since the top-quark contributions, that decouple in this limit,
are not included in the MCFM calculation of the 1-loop amplitudes.},
with timing results shown in Table~\ref{timings}. Although the evaluation of this part
of the full NNLO result is not the most expensive in terms of computing
time, our results are in all cases faster than both Openloops~2 and Recola2.
\begin{table}
\begin{center}
\begin{tabular}{|l|l| c|c|c|c|}
\hline 
Parton channel & Process & $\kappa$(OpenLoops 2) & $\kappa$(Recola2) & $t_{MCFM}$[s/1000 pts] \\ 
\hline 
$d \bar u \to e^- \bar\nu_e \gamma g$        & $W^-\gamma~+$~jet & 31.2 & 23.7 & $0.14$ \\
$u \bar d \to e^+ \nu_e \gamma g$            & $W^+\gamma~+$~jet & 29.1 & 24.3 & $0.14$ \\
$u \bar d \to e^+ e^- \gamma g$              & $Z \gamma~+$~jet  & 24.1 & 15.5 & $0.78$ \\
$u \bar u \to e^- \bar\nu_e \mu^+ \nu_\mu g$ & $W^+W^-~+$~jet    & 17.9 & 12.0 & $0.4 $ \\
$d \bar u \to e^- \bar\nu_e \mu^+ \mu^- g$   & $W^-Z~+$~jet      & 7.2  & 5.2  & $0.83$ \\
$u \bar d \to e^+ \nu_e \mu^+ \mu^- g$       & $W^+Z~+$~jet      & 7.1  & 5.2  & $0.83$ \\
$u \bar u \to e^- e^+ \mu^+ \mu^- g$         & $ZZ~+$~jet        & 15.8 & 3.8  & $3.6$ \\
\hline\end{tabular}
\caption{The relative timing of the OpenLoops 2 and Recola2 libraries, to the analytic
1-loop calculations in MCFM, for the calculation of a single partonic channel for each diboson process.
The speed-up factor when using MCFM rather than a library $X$ is denoted by $\kappa(X)$,
where $\kappa(X) = t_{X}/t_{MCFM}$ and the timings $t$ 
are established by computing results for 1000 phase-space points on
an Intel Xeon  E5-2650 2.60GHz CPU.}
\label{timings}
\end{center}
\end{table}  

\section{Comparative study of jettiness and $q_T$ slicing}
In this section we exploit the leading logarithmic dependence on the transverse
momentum cut, $q_T^{cut}$ and jettiness cut, $\tau^{cut}$, to define the appropriate variables
to compare the two approaches. The leading logarithmic behaviour of a colour singlet cross section
integrated up to a small cutoff value, $q_T^{cut}$, is
\begin{equation}
  \Sigma_T = \sigma_0 \exp\Big[-\frac{\alpha_s C_F}{2 \pi} \ln^2 ((q_T^{cut})^2/Q^2)\Big]= \sigma_0
  \exp\Big[-\frac{2\alpha_s C_F}{\pi} \ln^2 (q_T^{cut}/Q)\Big]\,,
\label{eq:SigmaT}
\end{equation}
where $\sigma_0$ is the Born level cross section. 
The corresponding leading log formula for zero-jettiness integrated up to a cut of value $\tau^{cut}$ is,
\begin{equation}
\Sigma_\tau = \sigma_0 \exp\Big[-\frac{\alpha_s C_F}{\pi} \ln^2 \frac{\tau^{cut}}{Q}\Big].
\label{eq:Sigmatau}
\end{equation}
A simple derivation of these two formulas at order $\alpha_s$ is given in Appendix~\ref{leadinglog}.

The resources needed for a computation of a given
accuracy is dominated by the calculation of the above-cut contribution.
Comparing Eqs.~(\ref{eq:SigmaT}) and~(\ref{eq:Sigmatau}) one therefore
expects a similar size for the contribution coming from the above cut region when the values of $\tau^{\rm{cut}}$ and $q_T^{\rm{cut}}$
are related by~\cite{Berger:2010xi},
\begin{equation}
\label{eq:tauequivQT}
\frac{\tau^{\rm{cut}}}{Q} \simeq \left(  \frac{q_T^{\rm{cut}}}{Q} \right)^{\sqrt{2}} \,.
\end{equation}
We therefore define the following two dimensionless quantities to encapsulate the slicing dependence of the results,
\begin{equation}
\epsilon_{T}=q_T^{\rm{cut}}/Q \,,
\label{eq:epsilonT}
\end{equation}
and
\begin{equation}
\epsilon_{\tau}= (\tau^{\rm{cut}}/Q)^{\frac{1}{\sqrt{2}}} \,.
\label{eq:epsilontau}
\end{equation}
The computational burden is then expected to be very similar for equal values
of $\epsilon_{T}$ and $\epsilon_{\tau}$ and therefore we will
compare the two schemes at the same values of $\epsilon_{T}$ and $\epsilon_{\tau}$.
Although this argument is only made at the level of leading logarithms, we will see
later (c.f. Table~\ref{nnlotimings} in Section~\ref{sec:nnlo}) that it is indeed supported
even at NNLO for the operating values of $\epsilon_\tau$ and $\epsilon_T$ that we choose.
We note that all the results presented in this paper are obtained using a modified
version of the MCFM-9.0 code, thus allowing the computation of cross sections at multiple
values of $\epsilon_{T}$ or $\epsilon_{\tau}$ in one run~\cite{Campbell:2019dru}.
In order to present a fair comparison between the two approaches
we generate the phase space in an identical way in both cases, one that has
not been optimized for either.

\subsection{Processes and cuts}
\label{sec:processes}

For simplicity, and in order to avoid issues associated with the application of
fiducial cuts in 2-body decays~\cite{Alekhin:2021xcu,Salam:2021tbm}, we present
results for inclusive $Z$, $W^\pm$, $H$, $ZH$ and $W^\pm H$ production.  Decays of the
$W$, $Z$ and $H$ bosons are not included and no cuts are applied.

The remaining processes we compute are:
\beqn
\label{processes_with_cuts}
pp &&\to \gamma\gamma \qquad \nonumber\\
pp &&\to e^- e^+ \gamma \qquad \quad\quad Z\gamma \nonumber\\
pp &&\to e^- \bar\nu_e \gamma \qquad \quad\quad W^-\gamma \nonumber\\
pp &&\to \nu_e e^+\gamma \qquad \quad\quad W^+\gamma \nonumber\\
pp &&\to e^- \mu^+ \bar\nu_e \nu_\mu \qquad\; W^-W^+ \nonumber\\ 
pp &&\to e^- e^+ \mu^- \mu^+ \qquad ZZ \nonumber\\ 
pp &&\to e^- \bar\nu_e \mu^- \mu^+ \qquad W^-Z \nonumber\\
pp &&\to \nu_e e^+ \mu^- \mu^+ \qquad W^+Z 
\eeqn
As indicated, these processes include a full set of contributing
Feynman diagrams and are not limited to those containing on-shell $W$
or $Z$ propagators.  We will, however, often use these names as
shorthand in the remainder of the paper.  The calculation of these
processes includes the application of cuts to identify photons and
leptons.  In order to provide an additional cross-check we have
adopted the sets of cuts used in Ref.~\cite{Grazzini:2017mhc} for
the processes in Eq.~(\ref{processes_with_cuts}).  The
cuts for the processes are given in Tables 7--10 of that reference.
For the convenience of the reader all the
cuts that we use for the various processes are given in
appendix~\ref{appendixcuts}.  We choose a common renormalization and
factorization scale $\mu$ that, however, depends upon the process as
follows:
$\mu = m_H$ ($H$),
$\mu = m_{V}$ ($V = W$ or $Z$),
$\mu = m_{VH}$ ($VH$),
$\mu = m_{\gamma\gamma}$ ($\gamma\gamma$),
$\mu = \sqrt{m_V^2 + (p_T^\gamma)^2}$ ($V\gamma$),
$\mu = (m_{V_1}+m_{V_2})/2$ ($V_1V_2$).

\subsection{Input parameters}
\begin{table}
\begin{center}
\begin{tabular}{|l|l|l|l|}
\hline
$M_W$ & 80.385~GeV                & $\Gamma_W$ & 2.0854~GeV\\
$M_Z$ & 91.1876~GeV               & $\Gamma_Z$ & 2.4952~GeV \\
$G_\mu$ & $1.166390\times10^{-5}$~GeV$^{-2}$   &   &\\
 $m_t$          & 173.2~GeV   &  $m_h$          & 125~GeV\\
\hline
\hline
\renewcommand{\baselinestretch}{1.8}
$m_W^2 = M_W^2-i M_W \Gamma_W $ & \multicolumn{3}{l|}{$(6461.748225 - 167.634879\, i) $~GeV$^2$} \\
$m_Z^2 = M_Z^2-i M_Z \Gamma_Z $ & \multicolumn{3}{l|}{$(8315.17839376 - 227.53129952\, i)$~GeV$^2$} \\
$\cos^2\theta_W={m_W^2}/{m_Z^2}$          & \multicolumn{3}{l|}{$(0.7770725897054007 + 0.001103218322282256\, i)$}\\
$\alpha = \frac{\sqrt{2}G_\mu}{\pi} M_W^2 (1-\frac{M_W^2}{M_Z^2})$ & \multicolumn{3}{l|}{$7.56246890198475\times10^{-3}$ giving $1/\alpha\approx 132.23\ldots$}\\
\hline
\end{tabular}
\caption{Input and derived parameters used for our numerical estimates.\label{parameters}}
\end{center}
\end{table}
\renewcommand{\baselinestretch}{1}

Most parameters are specified in Table~\ref{parameters}, where for generality we have
identified values in the complex mass scheme~\cite{Denner:2006ic}.  However,
for the calculation of the inclusive cross-sections (i.e. $Z$, $W^\pm$, $H$, $ZH$ and $W^\pm H$ production)
described in section~\ref{sec:processes}
all parameters are kept real by setting $\Gamma_Z = \Gamma_W = \Gamma_H = 0$.

For $W^\pm$ production we use a CKM matrix that employs the 2016 PDG
values~\cite{ParticleDataGroup:2016lqr}:
\begin{equation}
\left( \begin{array}{ccc}
 V_{ud}~ & V_{us}~ & V_{ub} \\
 V_{cd}~ & V_{cs}~ & V_{cb}
\end{array} \right) =
\left( \begin{array}{ccc}
 0.97417~ & 0.2248~ & 0.00409 \\
    0.22~ & 0.995 ~ & 0.0405
\end{array} \right)
\end{equation}
In all other processes we use a diagonal CKM matrix.  We use $n_f=5$ flavours of massless partons
throughout, except for the $W^-W^+$ process that uses $n_f=4$ to eliminate
contributions at NNLO from final states such as $W^-W^+b\bar b$
(that are considered a part of $t\bar t$ production and subsequent decay).
For the $W^\pm H$ and $ZH$ processes, diagrams in which the Higgs boson couples
directly to a top quark loop are computed in the effective theory that is valid in the large $m_t$ limit.
Contributions from a massive top quark are neglected in the
virtual 2-loop matrix elements for all processes, and also throughout the calculation of
NNLO corrections to the $W^-W^+$ and $Z\gamma$ processes.

All calculations are performed at $\sqrt{s}=13$~TeV and we use the NNPDF3.0 set of parton
distribution functions~\cite{NNPDF:2014otw}, with the set matched to the order of the
calculation and the number of quark flavours.

\subsection{NLO}
We first provide a set of illustrative results by computing results at NLO accuracy,
using both non-local slicing methods.  At
this order we can also compute benchmark results directly in MCFM 
using the subtraction method\cite{Ellis:1980wv} in the dipole formulation \cite{Catani:1996vz}.

\subsubsection{Inclusive processes}

\begin{figure}[t]
\includegraphics[width=0.35\textwidth,angle=270]{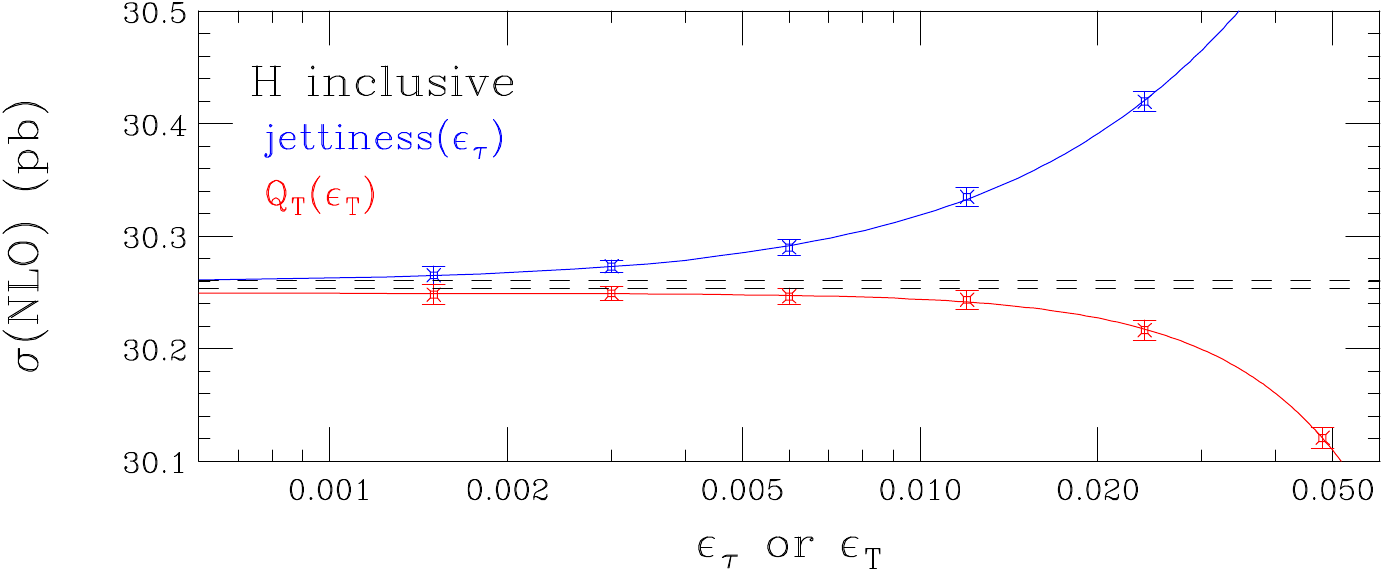}
\includegraphics[width=0.35\textwidth,angle=270]{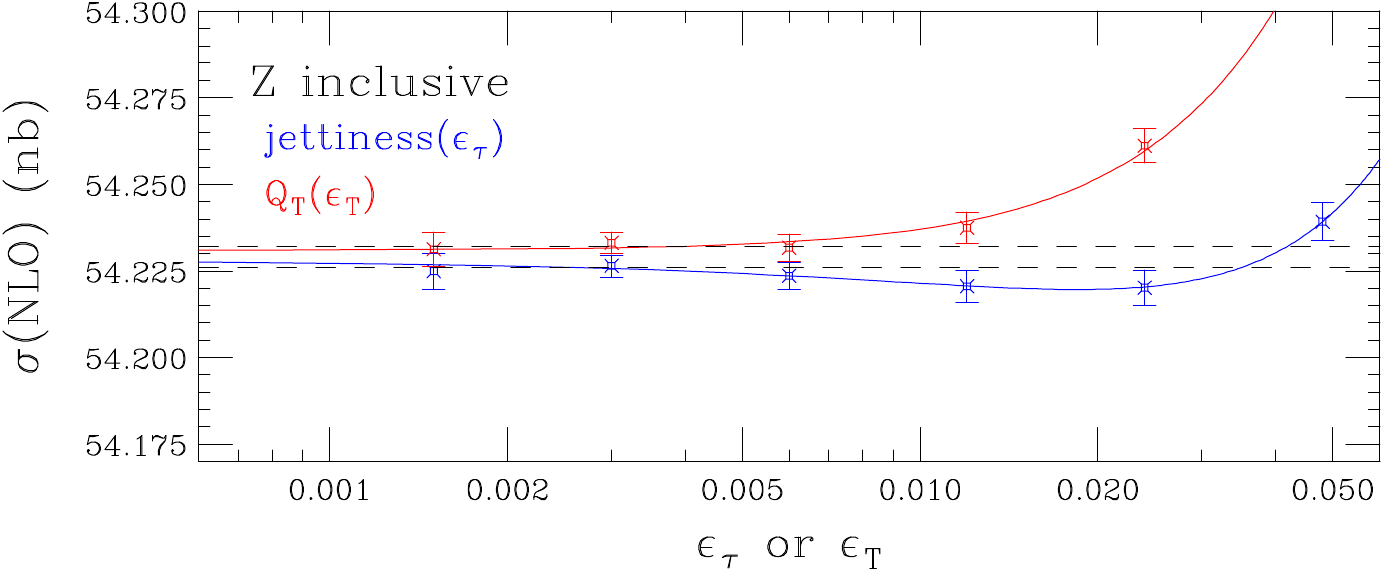}
\includegraphics[width=0.35\textwidth,angle=270]{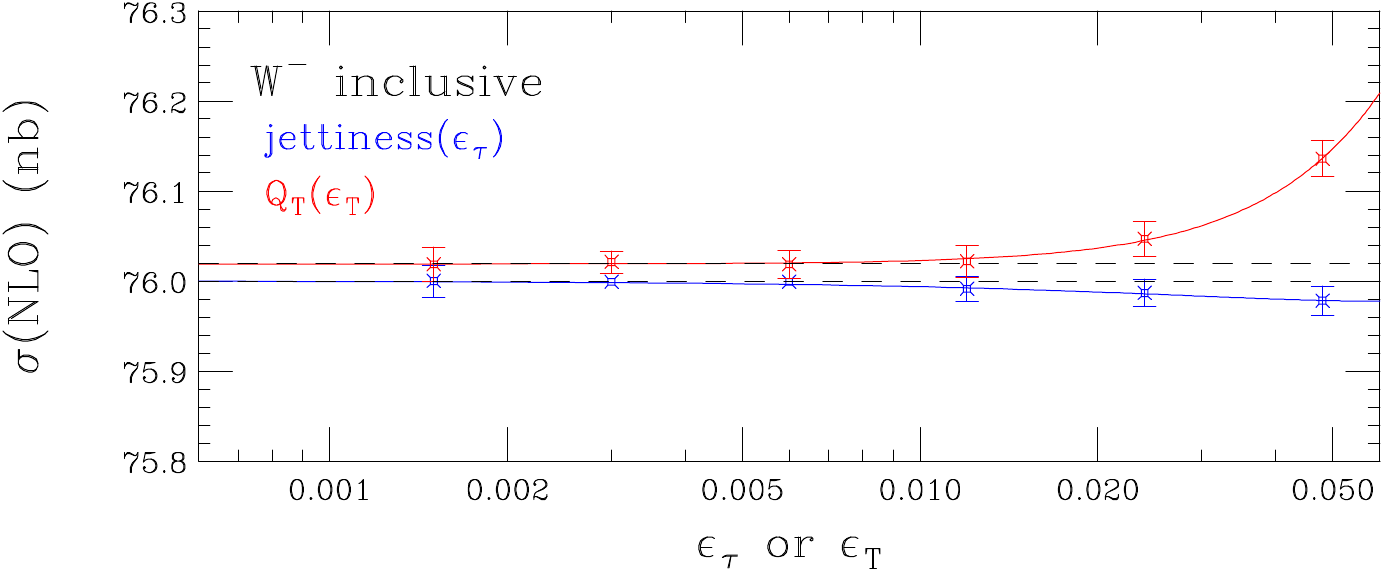}
\includegraphics[width=0.35\textwidth,angle=270]{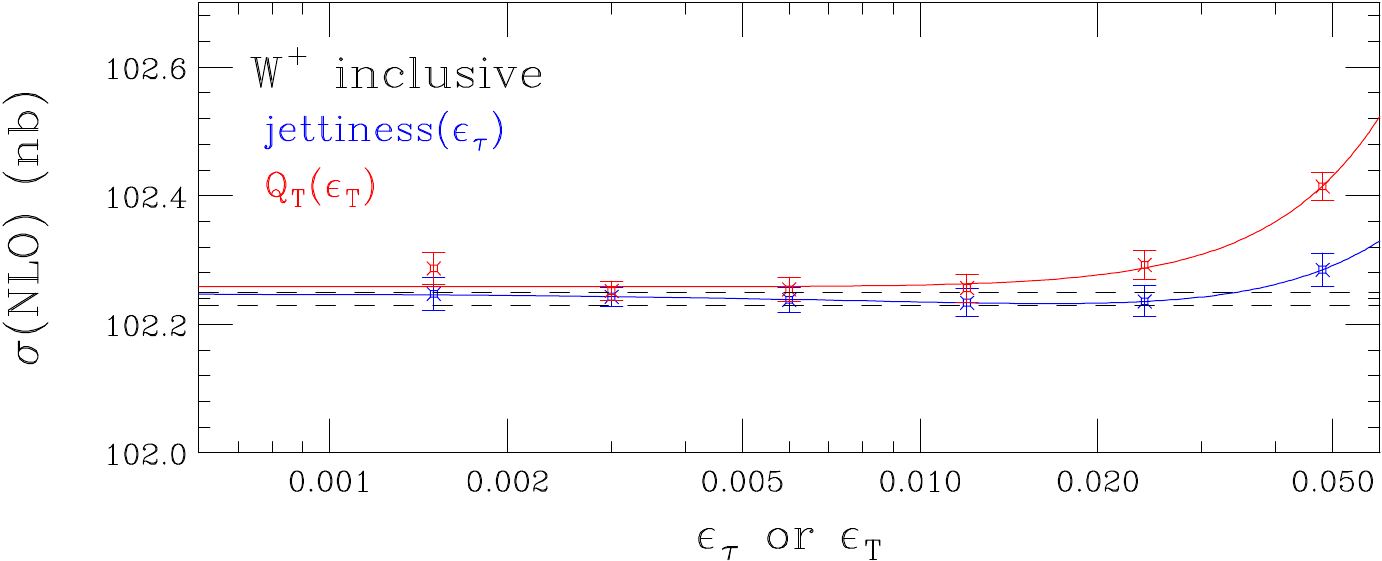}
\caption{Dependence of NLO cross section for $pp \to h$, $pp \to Z$,
$pp \to W^-$ and $pp \to W^+$ processes on choice of slicing cut, for both
0-jettiness and $q_T$-slicing.  The uncertainty band of the exact result, computed with MCFM using dipole subtraction,
is shown as the dashed lines.
\label{fig:2to1inclcutdepnlo}}
\end{figure}
\begin{figure}[t]
\includegraphics[width=0.35\textwidth,angle=270]{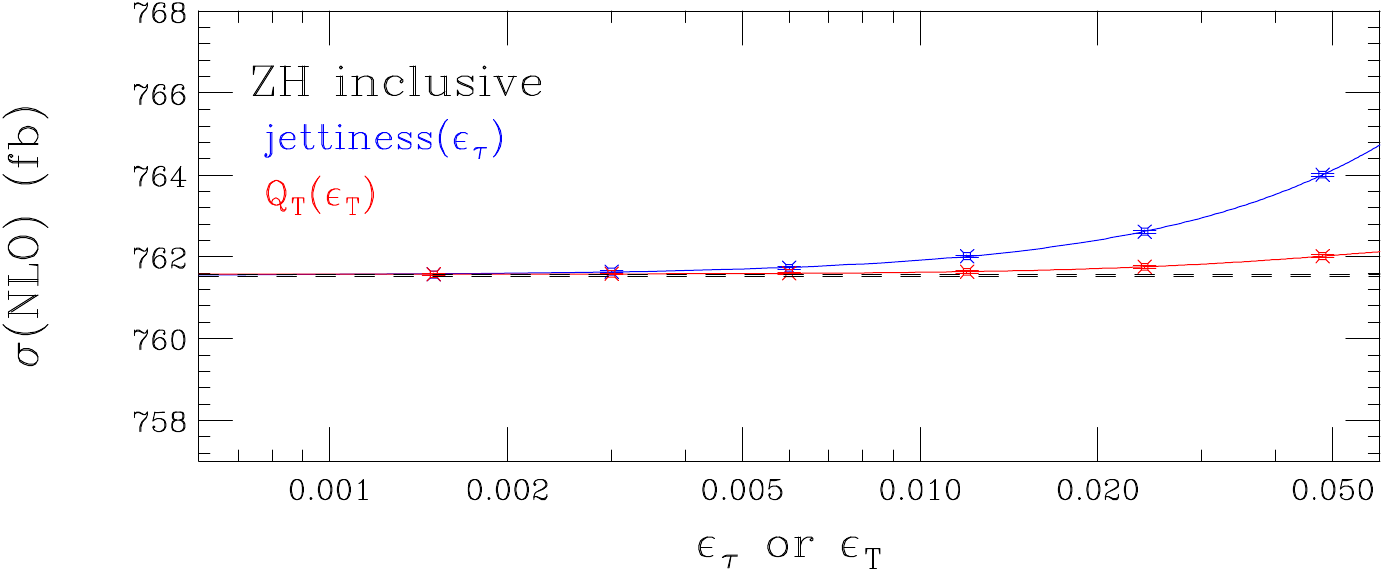}
\includegraphics[width=0.35\textwidth,angle=270]{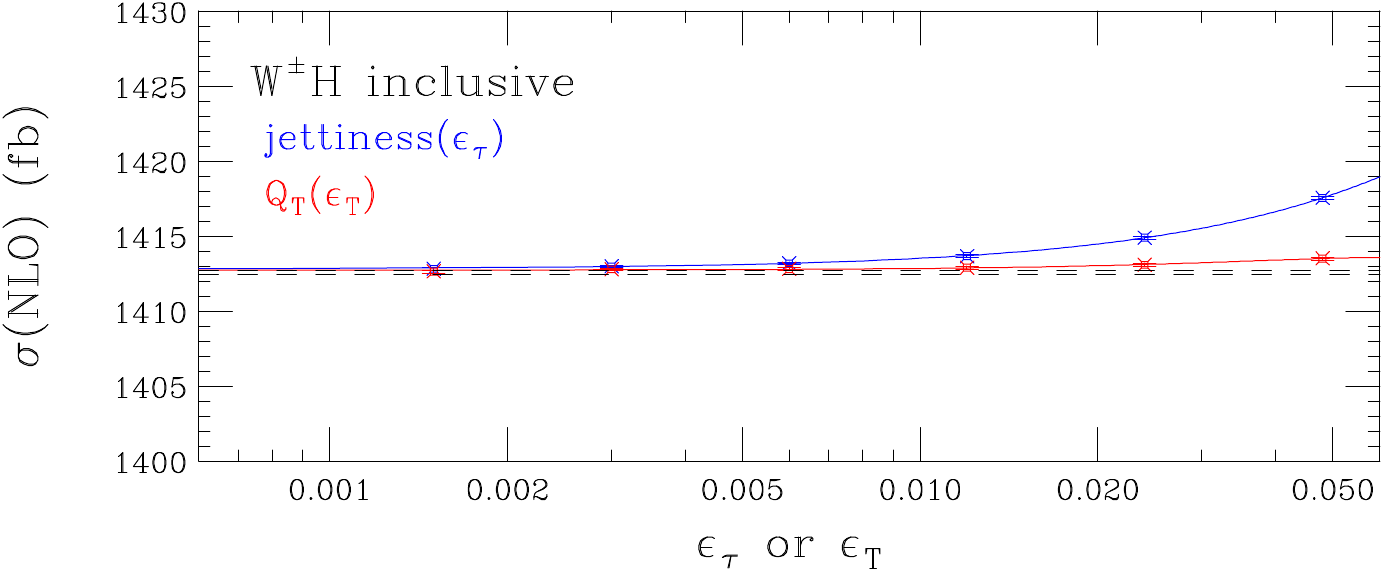}
\caption{Dependence of NLO cross section for inclusive $ZH$ and $W^\pm H$ (sum of $W^+H$ and $W^-H$) processes on choice of slicing cut, for both
0-jettiness and $q_T$-slicing.  The uncertainty band of the exact result, computed with MCFM using dipole subtraction,
is shown as the dashed lines.
\label{fig:2to2inclcutdepnlo}}
\end{figure}

The NLO results for inclusive production are shown in Fig.~\ref{fig:2to1inclcutdepnlo}
($2 \to 1$ processes) and Fig.~\ref{fig:2to2inclcutdepnlo} ($2 \to 2$ associated Higgs
production processes).  For $WH$ production we show the result for the sum of the two $W$ charges,
$\sigma(W^\pm H) = \sigma(W^+H) + \sigma(W^-H)$.
The result for the NLO cross section is computed as a
function of $\epsilon_{\tau}$ for 0-jettiness, c.f. Eq.~(\ref{eq:epsilontau}),
and as a function of $\epsilon_{T}$ for $q_T$-slicing, c.f. Eq.~(\ref{eq:epsilonT}).
Since the value of the cutoff used to present the NNLO results in Ref.~\cite{Grazzini:2017mhc}
is $\epsilon_{T}=q_T^{\rm{cut}}/Q = 0.15\%$ we perform the calculation at values of
$\epsilon$ with this as a lower bound.  We note that setting $\epsilon_{\tau}=0.15\%$
corresponds to $\tau^{\rm{cut}}/Q \approx 10^{-4}$, c.f. Eq.~(\ref{eq:epsilontau}).

The results from the non-local slicing schemes are
compared to those of MCFM dipole-subtraction calculations, which also
all agree fully with the results reported in Table 6 of
Ref.~\cite{Grazzini:2017mhc}.
A fit to the results at fixed values of $\epsilon_{T}$ and $\epsilon_{\tau}$
is performed using the  form,
\begin{equation}
\sigma^{NLO}(\epsilon) = a_0 + a_1 \epsilon^r \log\epsilon^r + a_2 \epsilon^r \,,
\label{eq:nlofitform}
\end{equation}
where $r=2$ ($q_T$) and $r=\sqrt2$ (0-jettiness) effectively undoes the rescaling
introduced in Eq.~(\ref{eq:epsilontau}).  This fit form anticipates the effect
of possible power corrections to the factorization theorems used in obtaining the
below-cut contribution (quadratic for $q_T$ and linear for 0-jettiness)
but here we only use this fit to guide the eye.

In all cases the results from the non-local slicing calculations approach the
known cross sections as the cutoff approaches zero.  For $H$, $W^\pm H$ and $ZH$ production
the residual difference from the known result is smaller for $q_T$ than 0-jettiness slicing,
for results at equal values of $\epsilon_{T}$ and $\epsilon_{\tau}$.  For the
$Z$ and $W^\pm$ processes the relative ordering is reversed.  We note that all the points
in these plots have been obtained by running the MCFM code with a target Monte Carlo precision
that is the same for 0-jettiness and $q_T$ slicing, so that 
the statistical errors on data points of equal $\epsilon_{T}$
and $\epsilon_{\tau}$ are similar.  The running time of the code to
reach this level of precision is essentially the same for the two non-local slicing methods, thereby
providing an indirect confirmation of the scaling behavior introduced in Eq.~(\ref{eq:tauequivQT}).

\subsubsection{Diboson production}

\begin{figure}[t]
\includegraphics[width=0.35\textwidth,angle=270]{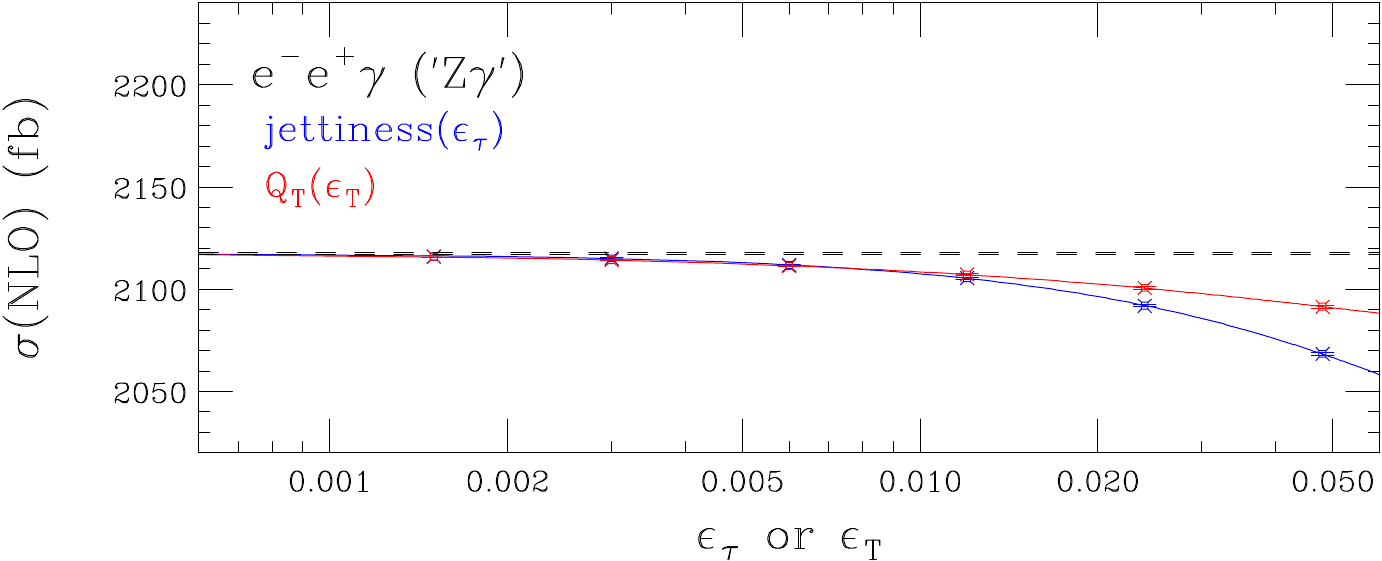}
\includegraphics[width=0.35\textwidth,angle=270]{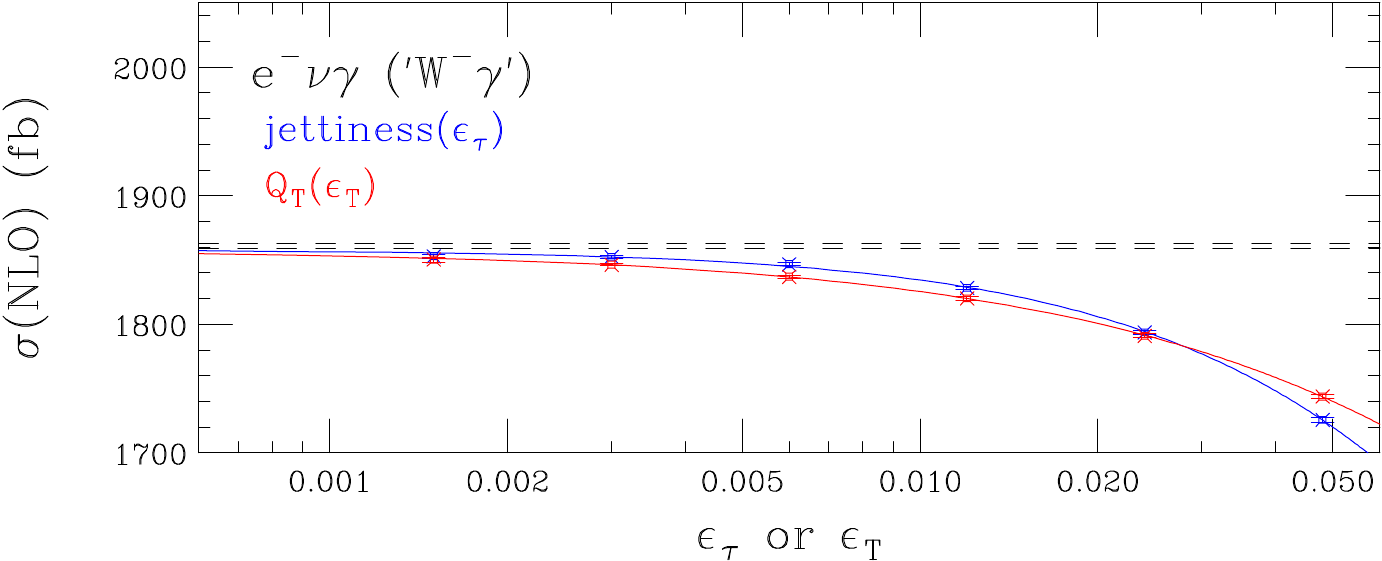}
\includegraphics[width=0.35\textwidth,angle=270]{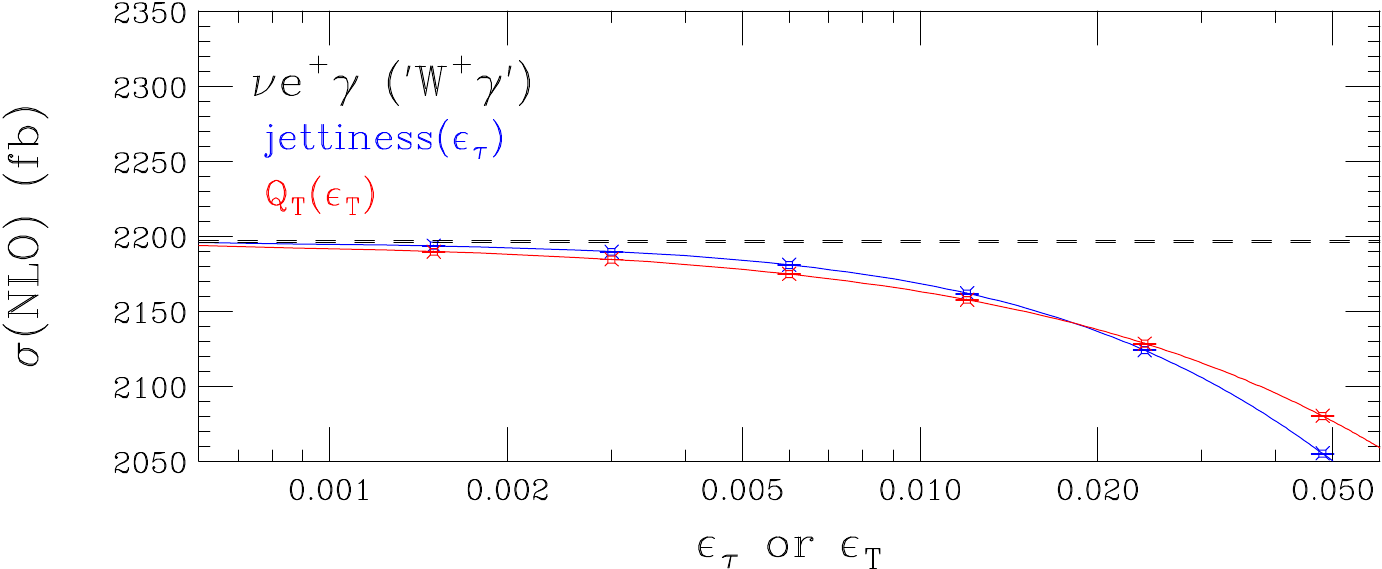}
\includegraphics[width=0.35\textwidth,angle=270]{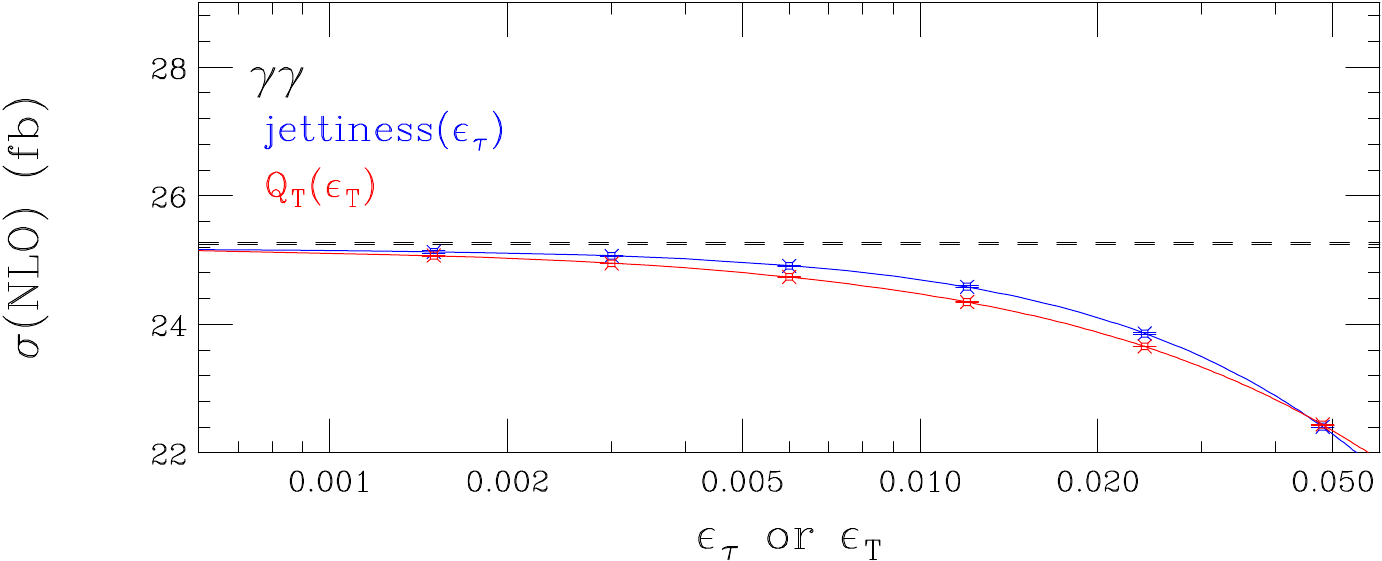}
\caption{Dependence of NLO cross section for $Z\gamma$, $W^-\gamma$, $W^+\gamma$ and
$\gamma\gamma$ processes on choice of slicing cut, for both
0-jettiness and $q_T$-slicing.  The uncertainty band of the exact result, computed with MCFM using dipole subtraction,
is shown as the dashed lines.
\label{fig:photoncutdepnlo}}
\end{figure}

\begin{figure}[t]
\includegraphics[width=0.35\textwidth,angle=270]{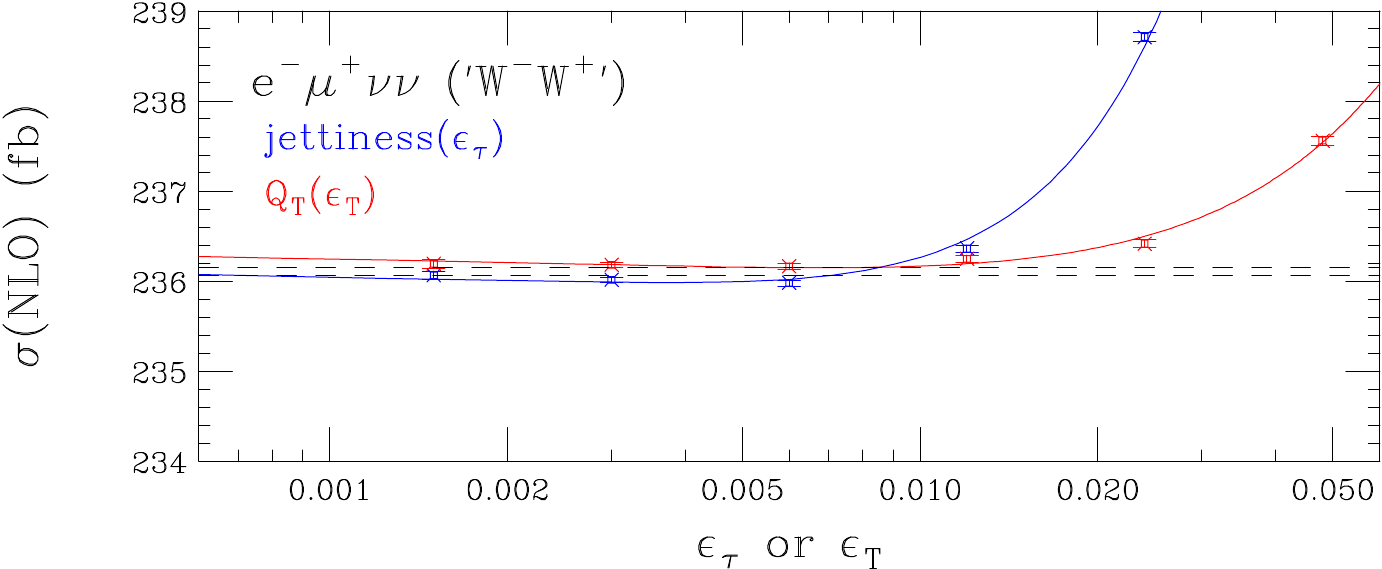}
\includegraphics[width=0.35\textwidth,angle=270]{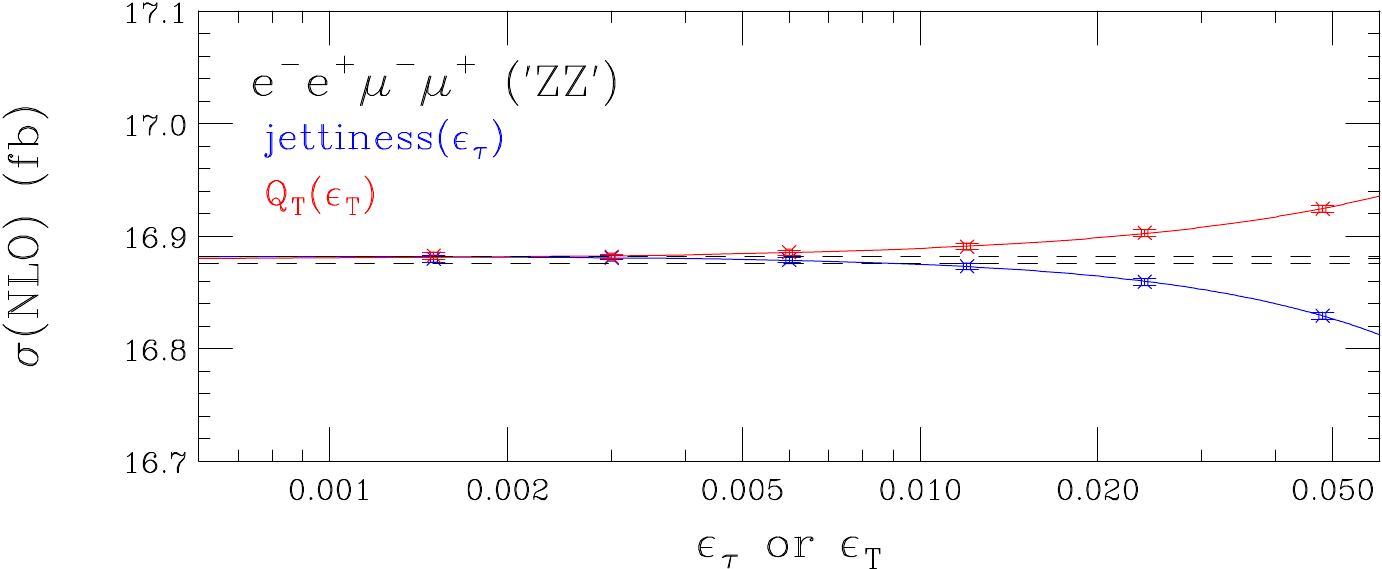}
\includegraphics[width=0.35\textwidth,angle=270]{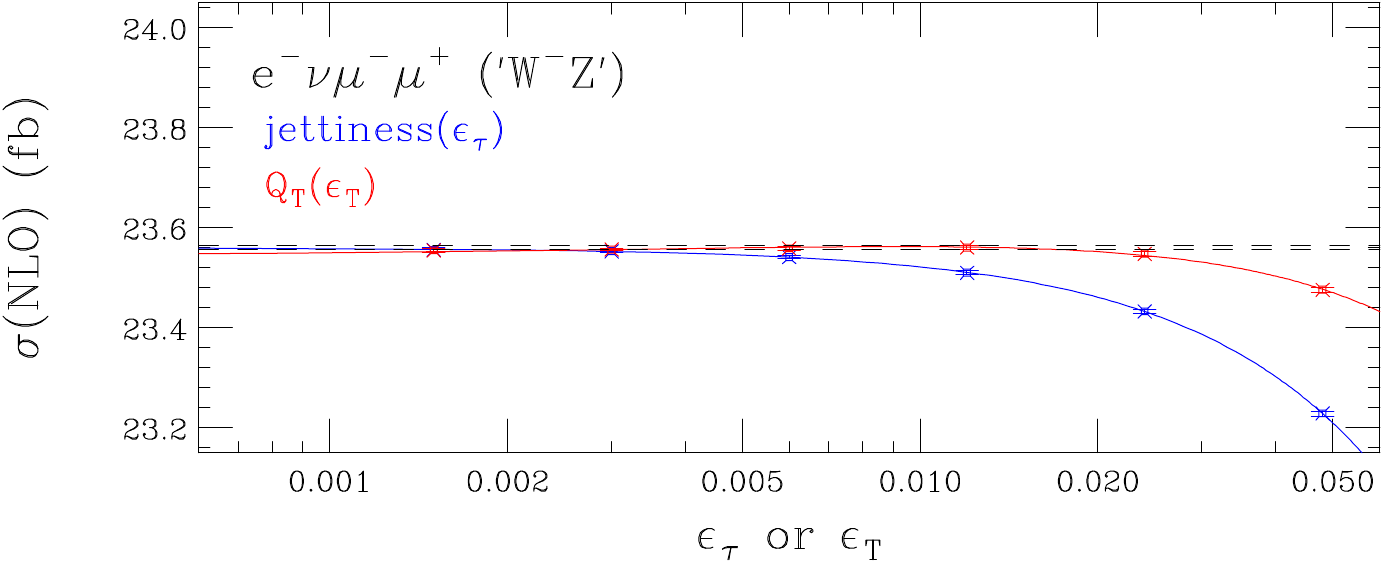}
\includegraphics[width=0.35\textwidth,angle=270]{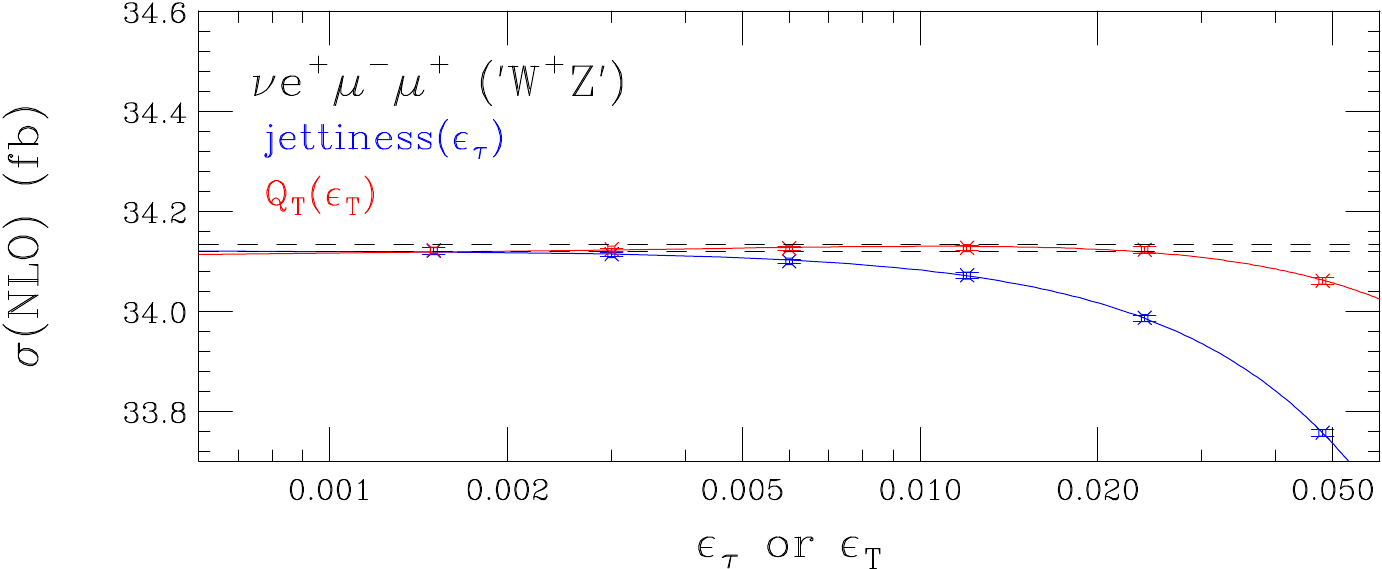}
\caption{Dependence of NLO cross section for $pp \to WW$, $pp \to ZZ$,
$pp \to W^-Z$ and $pp \to W^+Z$ processes on choice of slicing cut, for both
0-jettiness and $q_T$-slicing.  Dashed line is the NLO result computed with MCFM using dipole
subtraction.
\label{fig:dibosoncutdepnlo}}
\end{figure}

Corresponding results for processes involving a photon are shown in Fig.~\ref{fig:photoncutdepnlo}
and, for the remaining diboson cases, in Fig.~\ref{fig:dibosoncutdepnlo}. Note that here,
since the definition of these processes includes the application of fiducial cuts, we have fixed
$r=1$ in Eq.~(\ref{eq:nlofitform}) for $q_T$-slicing to anticipate the presence of linear power
corrections.  In~Fig.~\ref{fig:photoncutdepnlo}, processes in
which a final-state photon is observed, the approach to the known result is almost identical
for 0-jettiness and $q_T$ slicing.  This is also true for the $ZZ$ process, but for the
other diboson processes $q_T$ slicing is much closer to the correct result than 0-jettiness
for equal values of $\epsilon_{T}$ and $\epsilon_{\tau}$.

For the diphoton case we have also investigated the use of ``product cuts'', as advocated in
Ref.~\cite{Salam:2021tbm}, rather than the asymmetric cuts that are our default choice.  Since
we already observe no pathology in the asymmetric cut results of Fig.~\ref{fig:photoncutdepnlo}
the corresponding results for product cuts are qualitatively similar and we do not
show them separately here.  Since the study
of Ref.~\cite{Salam:2021tbm} is motivated by sensitivity specifically arising from the 2-body
decay of a parent particle this is expected;  the rapidly falling $p_T$ spectrum in the
continuum $pp \to \gamma\gamma$ case mitigates any similar issue here.

\clearpage

\subsection{NNLO}
\label{sec:nnlo}

Having established the format and pattern of results at NLO, we now turn our attention to NNLO.
At this order we may compare with the results of Ref.~\cite{Grazzini:2017mhc} for most processes
and for the remaining $W^\pm H$ and $ZH$ processes with the code {\tt vh@nnlo}~\cite{Brein:2003wg,Brein:2012ne}.
To focus more closely on the behaviour of the calculation at this order we will show results
not for the total NNLO cross section, but for the ${\cal O}(\alpha_s^2)$ contribution that enters
at this order.  In order to extract benchmark predictions for this quantity from
Refs.~\cite{Grazzini:2017mhc,Brein:2003wg,Brein:2012ne} we have computed the NLO cross section for each process using
NNLO PDFs (denoted $\sigma_{NLO^*}$ in the table)
and subtracted these values from the corresponding NNLO results.  To compare with the MATRIX cross sections this method
is used to obtain results for $\epsilon_T = 0.15\%$ and after their extrapolation procedure.
The target NNLO corrections for comparison purposes are shown in Table~\ref{tab:targets}, as well
as the results of the NNLO calculations using MCFM. 

\begin{table}
\begin{center}
\begin{tabular}{l|r|rr|rrl}
\textbf{Process} 		  	   &                   & target       &                    & MCFM  \\
					   & $\sigma_{NLO^*}$  & $\sigma_{NNLO}$ & $\delta_{NNLO}$ & $\sigma_{NNLO}$ & $\delta_{NNLO}$ \\ 
\hline 
$pp\rightarrow H$			   & $29.78(0)$ &  39.93(3)  & $10.15(3)$  &  39.91(5)  & $10.13(5)$ & nb \\   
$pp\rightarrow Z$			   & $56.41(0)$ &  55.99(3)  & $-0.42(3)$  &  56.03(3)  & $-0.38(3)$ & nb \\   
$pp\rightarrow W^-$			   & $79.09(0)$ &  78.33(8)  & $-0.76(8)$  &  78.41(6)  & $-0.68(6)$ & nb \\   
$pp\rightarrow W^+$			   & $106.2(0)$ &  105.8(1)  & $-0.4(1)$   &  105.8(1)  & $-0.4(1)$ & nb \\
$pp\rightarrow \gamma\gamma$     	   & $25.61(0)$ &  40.28(30) & $14.67(30)$ &  40.19(20) & $14.58(20)$ & pb \\
$pp\rightarrow e^-e^+\gamma$      	   & $2194(0)$  &  2316(5)   & $122(5)$    &  2315(5)	& $121(5)$ & pb \\
$pp\rightarrow e^-\bar{\nu_e}\gamma$  	   & $1902(0)$  &  2256(15)  & $354(15)$   &  2251(2)   & $349(2)$ & pb \\
$pp\rightarrow e^+\nu_e\gamma$		   & $2242(0)$  &  2671(35)  & $429(35)$   &  2675(2)   & $433(2)$ & pb \\
$pp\rightarrow e^-\mu^-e^+\mu^+$	   & $17.29(0)$ & $20.30(1)$ & $3.01(1)$   & $20.30(2)$ & $3.01(2)$ & fb \\
$pp\rightarrow e^-\mu^+\nu_\mu\bar{\nu_e}$ & $243.7(1)$ & $264.6(2)$ & $20.9(3)$   & $264.9(9)$ & $21.2(8)$ & fb \\
$pp\rightarrow e^-\mu^-e^+\bar{\nu_\mu}$   & $23.94(1)$ & $26.17(2)$ & $2.23(3)$   & $26.18(3)$ & $2.24(2)$ & fb \\
$pp\rightarrow e^-e^+\mu^+\nu_\mu$	   & $34.62(1)$ & $37.74(4)$ & $3.12(5)$   & $37.78(4)$ & $3.16(3)$ & fb \\
$pp\rightarrow ZH$			   & $780.0(4)$ &  846.7(5)  & $66.7(6)$   &  847.3(7)  & $67.3(6)$ & fb \\    
$pp\rightarrow W^\pm H$			   & $1446.5(7)$&  1476.1(7) & $29.6(10)$  &  1476.7(8) & $30.2(4)$ & fb  \\
\end{tabular}
\caption{NLO results, computed using MCFM with NNLO PDFs (denoted $\sigma_{NLO^*}$), total NNLO cross sections
from {\tt vh@nnlo} ($W^\pm H$ and $ZH$ only) and MATRIX (remaining processes, using the extrapolated result
from Table 6 of Ref.~\cite{Grazzini:2017mhc}) and the target NNLO coefficients ($\delta_{NNLO}$,
with $\delta_{NNLO} = \sigma_{NNLO} - \sigma_{NLO^*}$). The result of the MCFM calculation
(0-jettiness, fit result $b_0$ from Eq.~(\ref{eq:nnlofit})) is shown in the final column.
\label{tab:targets}}
\end{center}
\end{table}

As at NLO, for each process we have used a target numerical precision for the calculation of each
process in order to compare the jettiness and $q_T$ slicing methods.  The actual precisions attained
and the corresponding CPU times required, for the operating points $\epsilon_{\tau}=0.15\%$ and
$\epsilon_{T}=0.15\%$, are shown in Table~\ref{nnlotimings}.  The time required for each
slicing method to reach a similar level of precision is very close for all processes, suggesting
that the scaling introduced in Eq.~(\ref{eq:epsilontau}) remains valid at NNLO for these values of
the slicing parameters.  The timings for the processes $pp\rightarrow e^-\bar{\nu_e}\gamma$ and
$pp\rightarrow e^+\nu_e\gamma$ differ the most, suggesting that the presence of an identified
photon in these proceses may alter the scaling somewhat (at least, under these cuts).  However,
the timings are still not dissimilar, especially given the computational effort that must be
employed for even small gains in numerical precision at this point ($\sim 0.3\%$ on the NNLO
coefficient). 

\begin{table}
\begin{center}
\begin{tabular}{ l|r|c|c}
\textbf{Process} & method & rel. unc. on $\delta_{NNLO}$ & time (CPU days) \\ \hline 
\multirow{2}{*}{$pp\rightarrow H$}			   & jettiness & 0.0029 & 54.7 \\
                                                           & $q_T$     & 0.0029 & 54.7 \\ \hline 
\multirow{2}{*}{$pp\rightarrow Z$}			   & jettiness & 0.045  & 356 \\
                                                           & $q_T$     & 0.039  & 364 \\ \hline 
\multirow{2}{*}{$pp\rightarrow W^-$}			   & jettiness & 0.029  & 274 \\
                                                           & $q_T$     & 0.029  & 277 \\ \hline 
\multirow{2}{*}{$pp\rightarrow W^+$}			   & jettiness & 0.084	& 238 \\
                                                           & $q_T$     & 0.086  & 275 \\ \hline 
\multirow{2}{*}{$pp\rightarrow \gamma\gamma$}     	   & jettiness & 0.0090 & 0.77 \\
                                                           & $q_T$     & 0.0079 & 0.89 \\ \hline 
\multirow{2}{*}{$pp\rightarrow e^-e^+\gamma$}      	   & jettiness & 0.023  & 340 \\
                                                           & $q_T$     & 0.024  & 330 \\ \hline 
\multirow{2}{*}{$pp\rightarrow e^-\bar{\nu_e}\gamma$}  	   & jettiness & 0.0032 & 310 \\
                                                           & $q_T$     & 0.0029 & 220 \\ \hline 
\multirow{2}{*}{$pp\rightarrow e^+\nu_e\gamma$}		   & jettiness & 0.0029 & 317 \\
                                                           & $q_T$     & 0.0028 & 231 \\ \hline 
\multirow{2}{*}{$pp\rightarrow e^-\mu^-e^+\mu^+$}	   & jettiness & 0.0040 & 317 \\
                                                           & $q_T$     & 0.0039 & 358 \\ \hline 
\multirow{2}{*}{$pp\rightarrow e^-\mu^+\nu_\mu\bar{\nu_e}$}& jettiness & 0.012  & 431 \\
                                                           & $q_T$     & 0.013  & 395 \\ \hline 
\multirow{2}{*}{$pp\rightarrow e^-\mu^-e^+\bar{\nu_\mu}$}  & jettiness & 0.0046 & 343 \\
                                                           & $q_T$     & 0.0053 & 323 \\ \hline 
\multirow{2}{*}{$pp\rightarrow e^-e^+\mu^+\nu_\mu$}	   & jettiness & 0.0048 & 441 \\
                                                           & $q_T$     & 0.0052 & 359 \\ \hline 
\multirow{2}{*}{$pp\rightarrow ZH$}			   & jettiness & 0.0047 & 87.3 \\
                                                           & $q_T$     & 0.0046 & 89.9 \\ \hline 
\multirow{2}{*}{$pp\rightarrow W^\pm H$}                   & jettiness & 0.021  & 47.5 \\
                                                           & $q_T$     & 0.019  & 46.8 \\ \hline
\end{tabular}
\caption{Summary of run parameters for the NNLO calculations presented in this paper.  For each process,
the table indicates the relative uncertainty on the NNLO coefficient ($\delta_{NNLO}$) that is computed
(for the lowest values of $\epsilon_\tau$ and $\epsilon_T$ shown in this paper), as well as the time
taken (in CPU days) to perform each calculation.
\label{nnlotimings}}
\end{center}
\end{table}

\subsubsection{Inclusive production}

\begin{figure}[t]
\includegraphics[width=0.35\textwidth,angle=270]{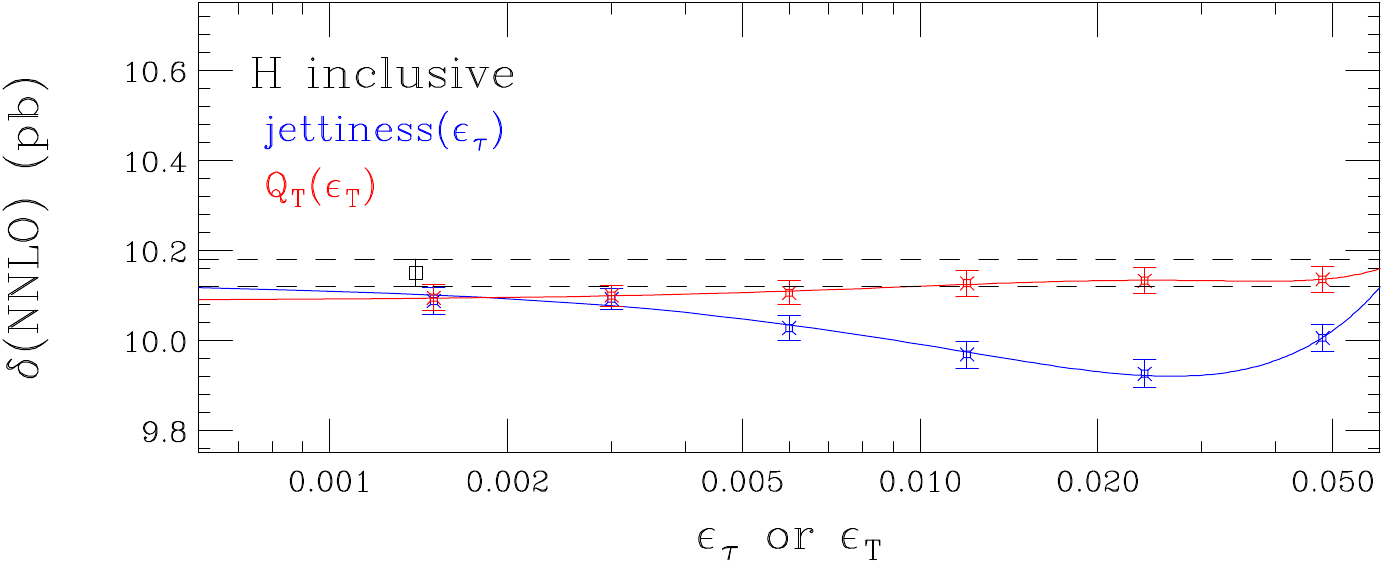}
\includegraphics[width=0.35\textwidth,angle=270]{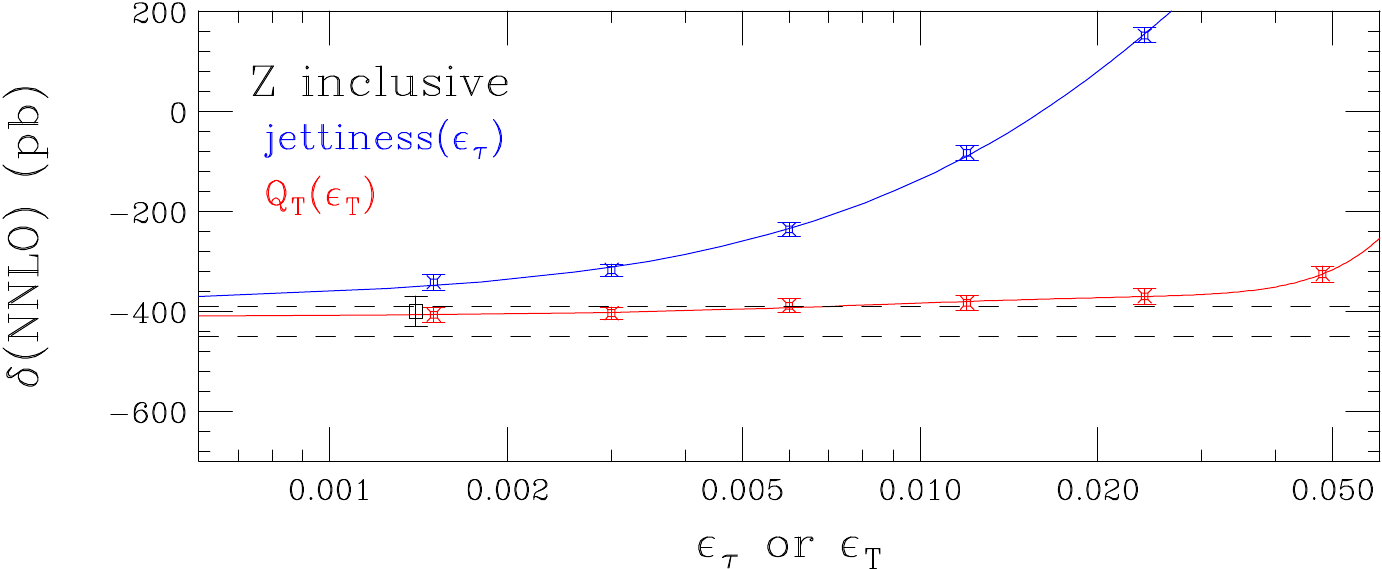}
\includegraphics[width=0.35\textwidth,angle=270]{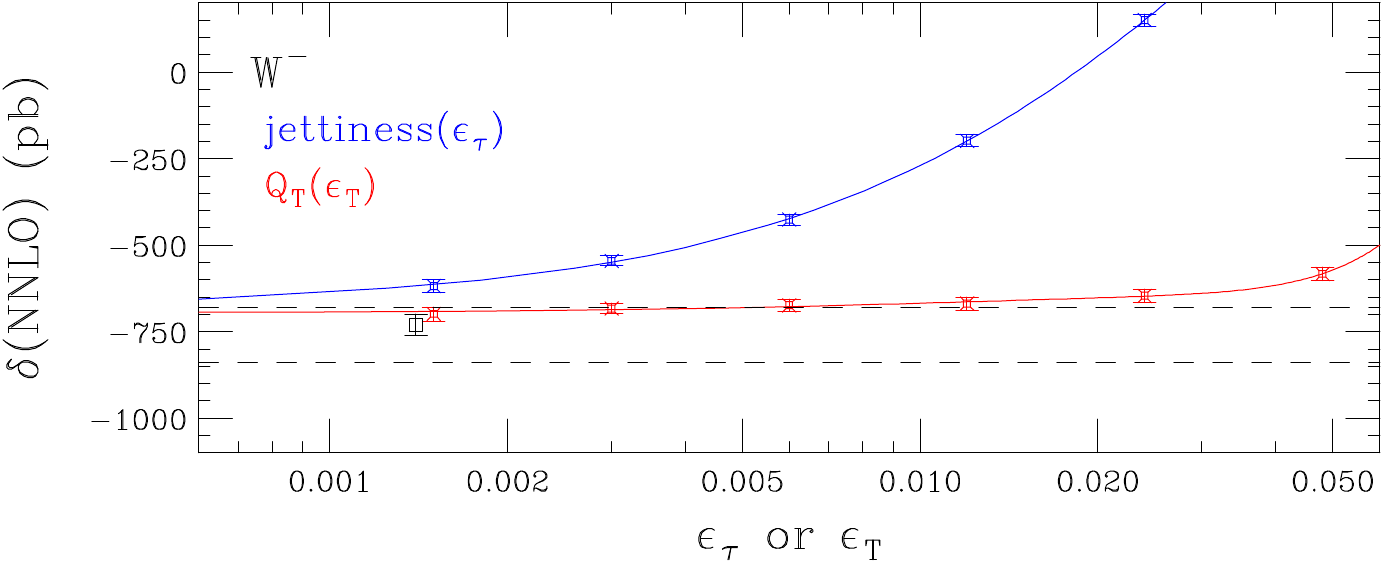}
\includegraphics[width=0.35\textwidth,angle=270]{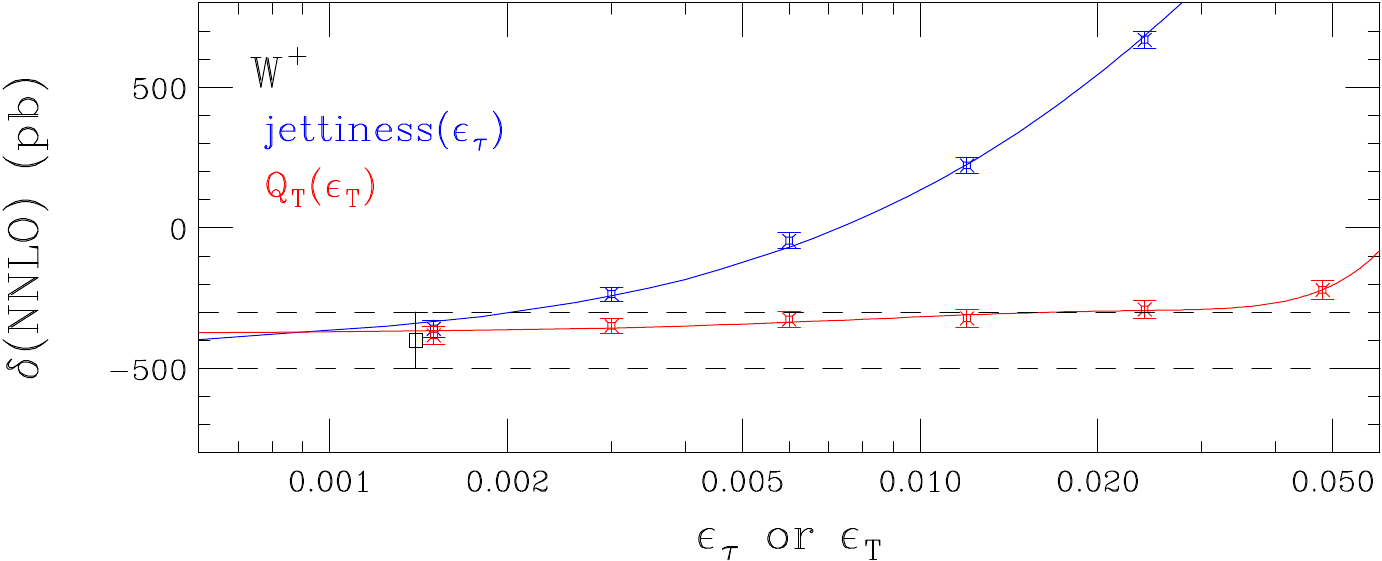}
\caption{Dependence of NNLO coefficient for inclusive $H$, $Z$, $W^-$ and $W^+$ processes on choice of slicing cut, for both
0-jettiness and $q_T$-slicing.
The MATRIX result for $q_T^{\rm cut} = 0.15\%$, Ref.~\cite{Grazzini:2017mhc} corresponds to the square black point (slightly offset
for visibility) and the uncertainty band of the extrapolated MATRIX result is shown as the dashed lines.
\label{fig:2to1inclcutdep}}
\end{figure}

\begin{figure}[t]
\includegraphics[width=0.35\textwidth,angle=270]{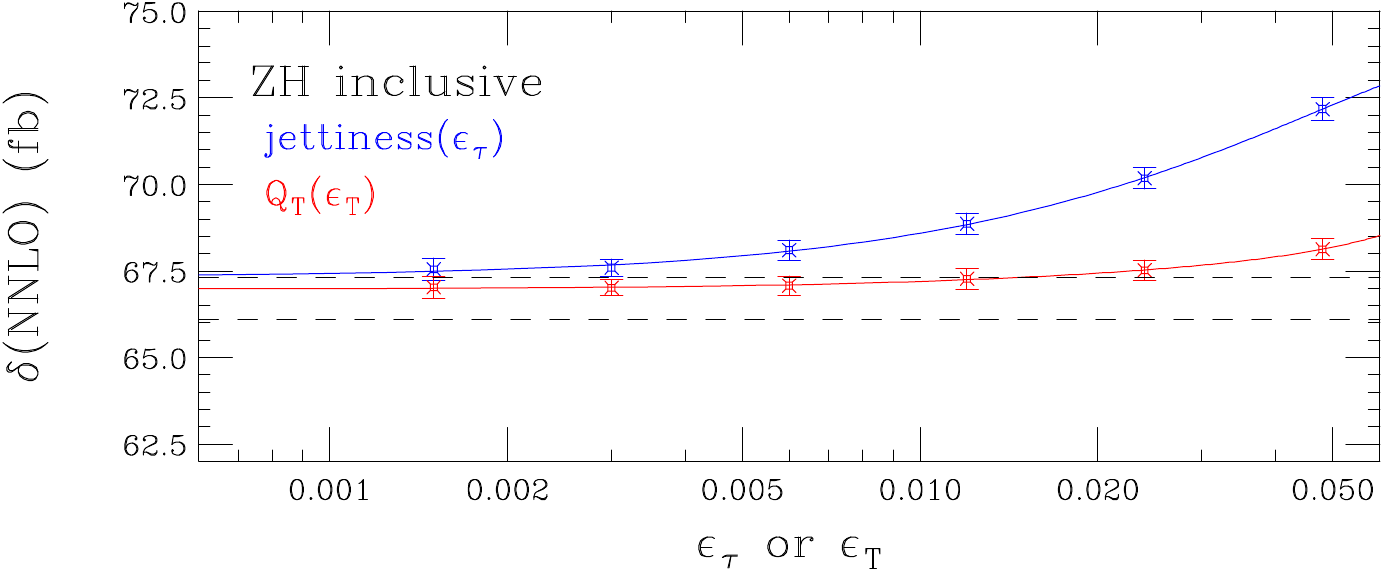}
\includegraphics[width=0.35\textwidth,angle=270]{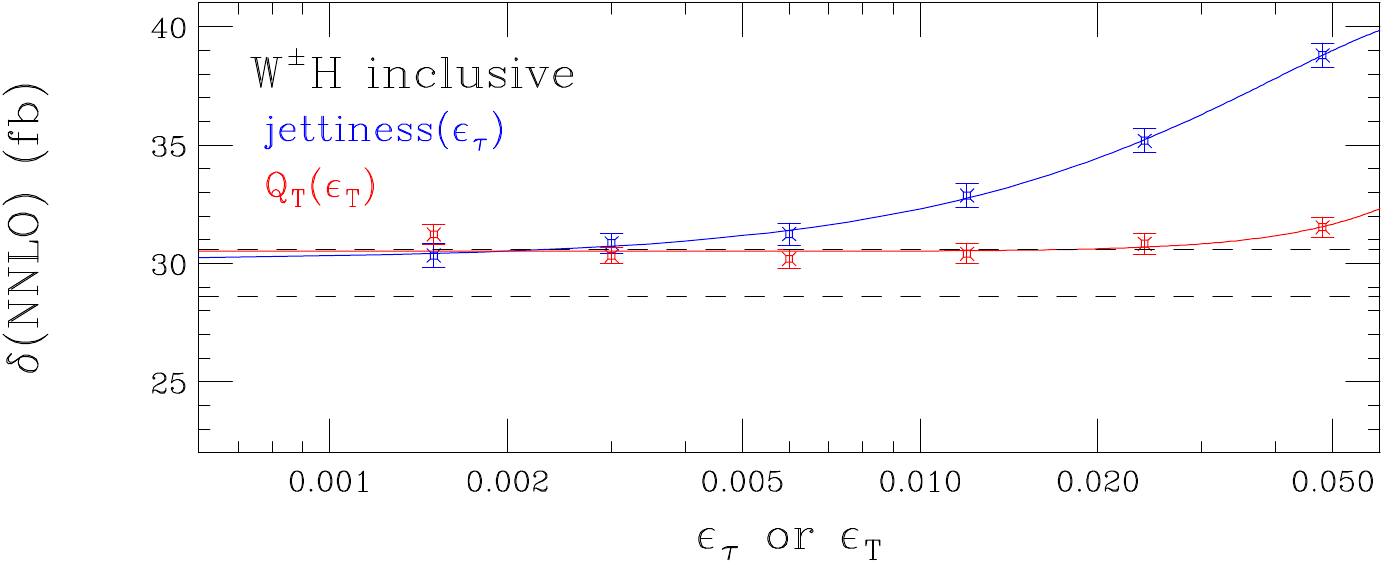}
\caption{Dependence of NNLO coefficient for inclusive $ZH$ and $W^\pm H$ (sum of $W^+H$ and $W^-H$) 
processes on choice of slicing cut, for both
0-jettiness and $q_T$-slicing.  The dashed lines represent the uncertainty band of the
{\tt vh@nnlo} result~\cite{Brein:2003wg,Brein:2012ne}.
\label{fig:2to2inclcutdep}}
\end{figure}

Results for the NNLO corrections to the inclusive calculations considered in this
paper are shown in Figs.~\ref{fig:2to1inclcutdep} and~\ref{fig:2to2inclcutdep}.
As at NLO we also show a fit to the data points, but this time using a form 
appropriate for power corrections that could be present at NNLO,
\begin{equation}
\sigma^{NNLO}(\epsilon) = b_0 + b_1 \epsilon^r \log^3\epsilon^r
 + b_2 \epsilon^r \log^2\epsilon^r + b_3 \epsilon^r \,.
\label{eq:nnlofit}
\end{equation}
For the $2\to1$ processes shown in Fig.~\ref{fig:2to1inclcutdep} we also indicate
the MATRIX result for $\epsilon_T =0.15\%$ and the extrapolated result,
as given in table~\ref{tab:targets}, from the same calculation.
For the associated Higgs production processes we also show the {\tt vh@nnlo} results
in Fig.~\ref{fig:2to2inclcutdep}.

The results from the two non-local subtraction schemes are in excellent agreement
in the limit $\epsilon \to 0$, and in the case of the $2\to1$ processes, also match
those extracted from Ref.~\cite{Grazzini:2017mhc}.  The approach to this limit differs
substantially between the two subtraction schemes;  results in the $q_T$ scheme
are much closer to the asymptotic value across the range while, in contrast, 
$0$-jettiness suffers from much larger corrections at finite values of
$\epsilon_{\tau}$. 

\subsubsection{Diboson production}
\begin{figure}[t]
\includegraphics[width=0.35\textwidth,angle=270]{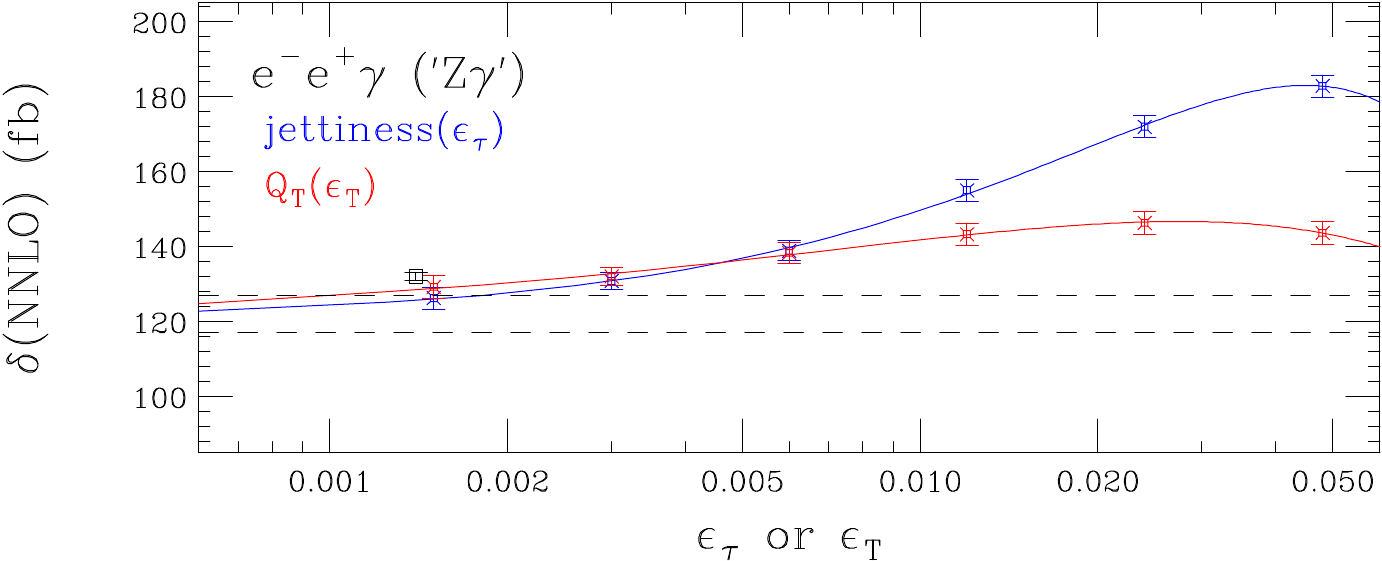}
\includegraphics[width=0.35\textwidth,angle=270]{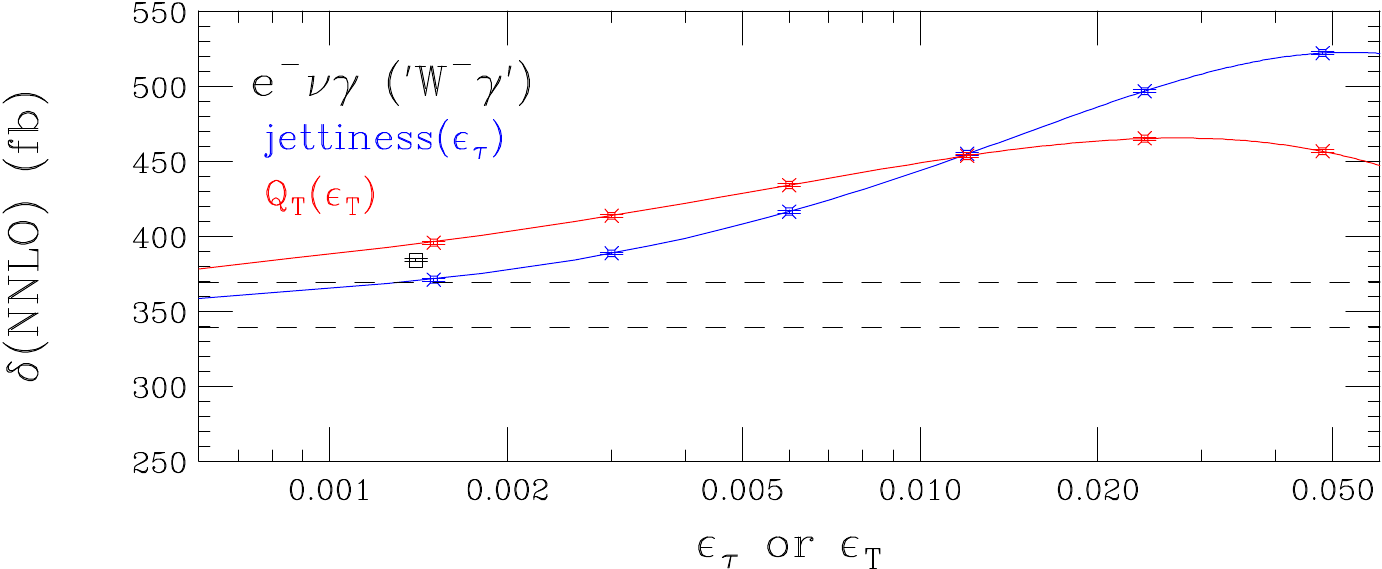}
\includegraphics[width=0.35\textwidth,angle=270]{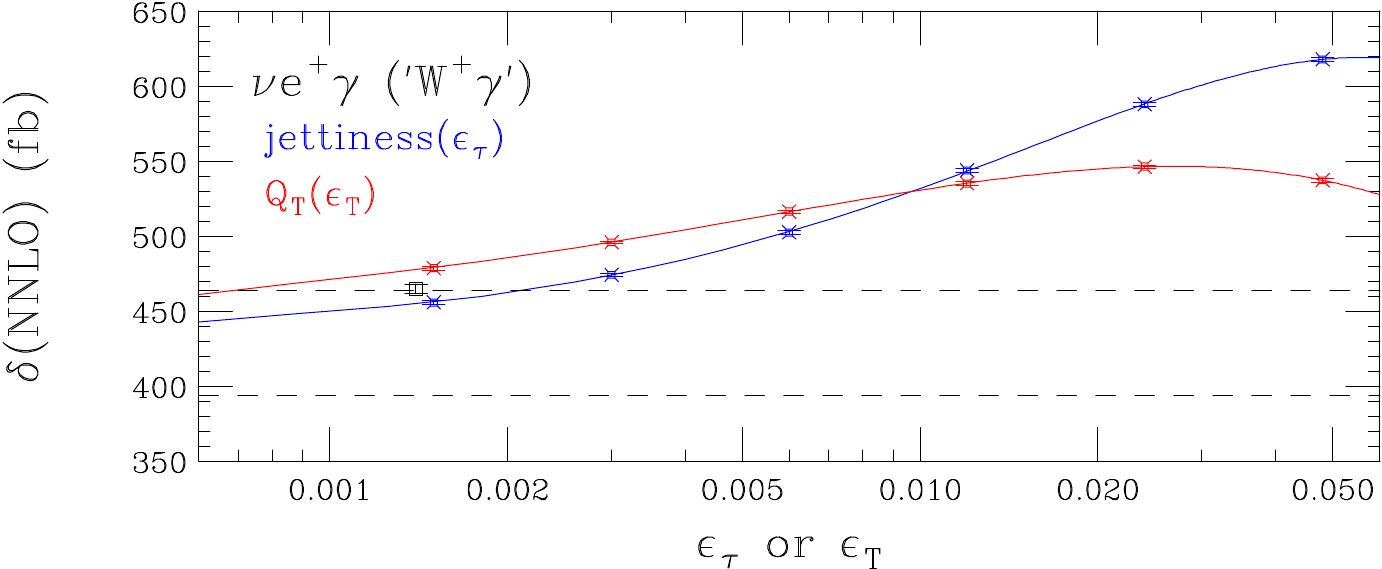}
\includegraphics[width=0.35\textwidth,angle=270]{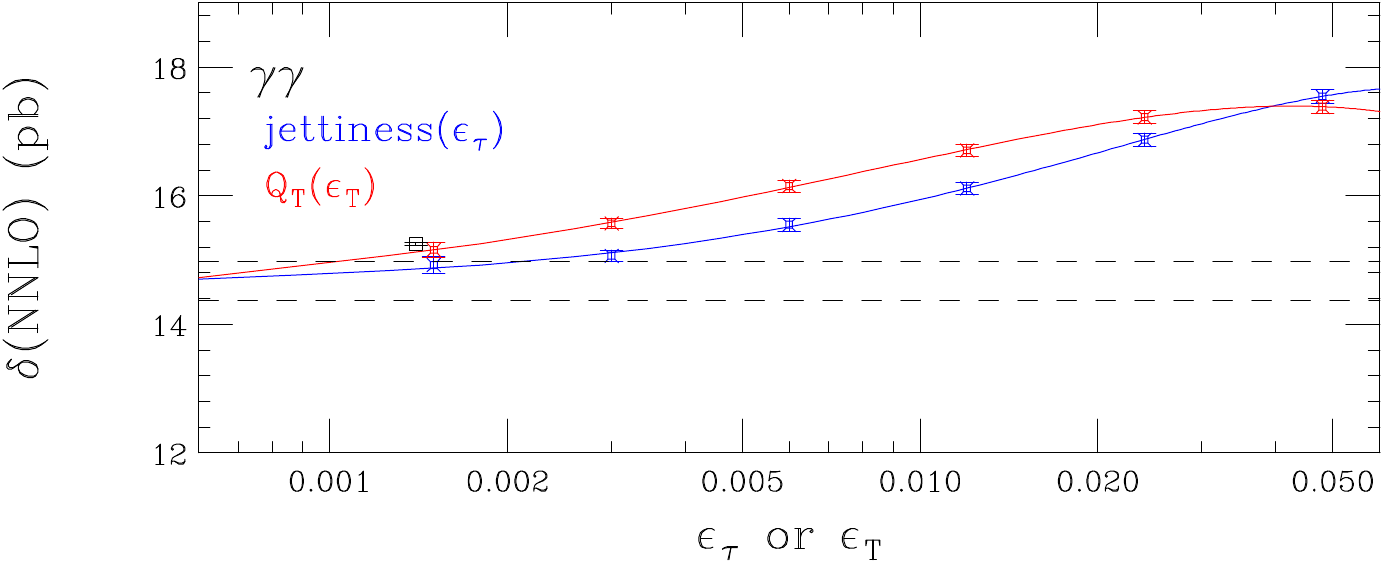}
\caption{Dependence of NNLO coefficient for $Z\gamma$, $W^-\gamma$, $W^+\gamma$ and
$\gamma\gamma$ processes on choice of slicing cut, for both
0-jettiness and $q_T$-slicing. 
The MATRIX result for $q_T^{\rm cut} = 0.15\%$, Ref.~\cite{Grazzini:2017mhc} corresponds to the square black point (slightly offset
for visibility) and the uncertainty band of the extrapolated MATRIX result is shown as the dashed lines.
\label{fig:photoncutdep}}
\end{figure}

\begin{figure}[t]
\includegraphics[width=0.35\textwidth,angle=270]{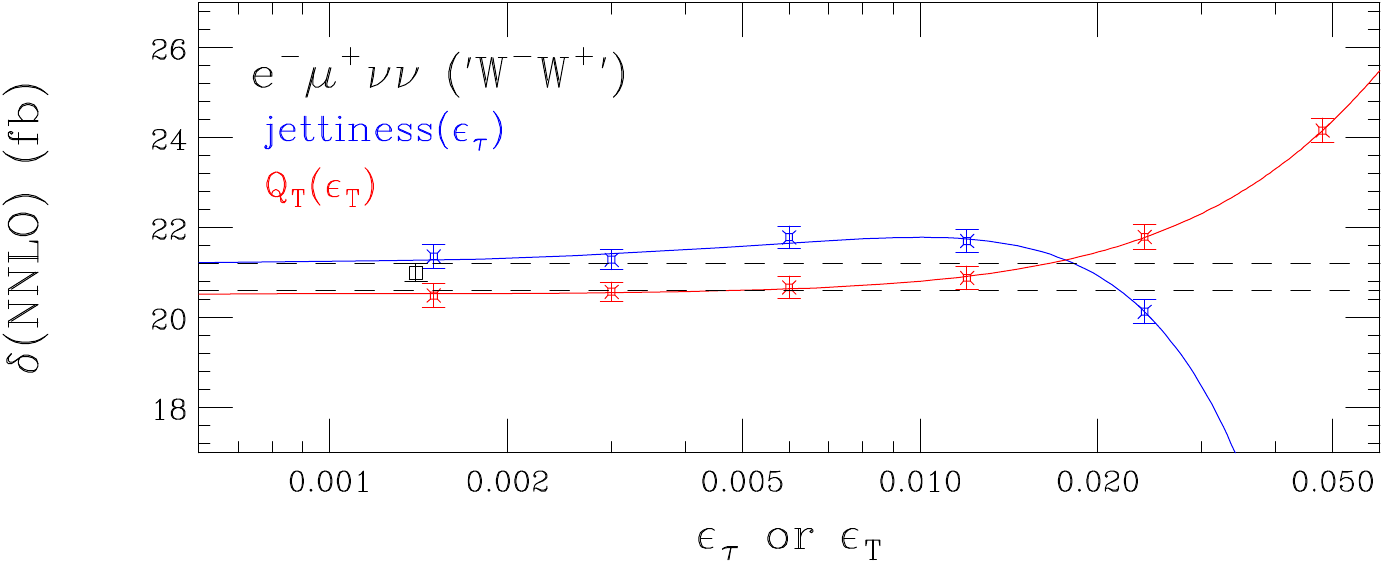}
\includegraphics[width=0.35\textwidth,angle=270]{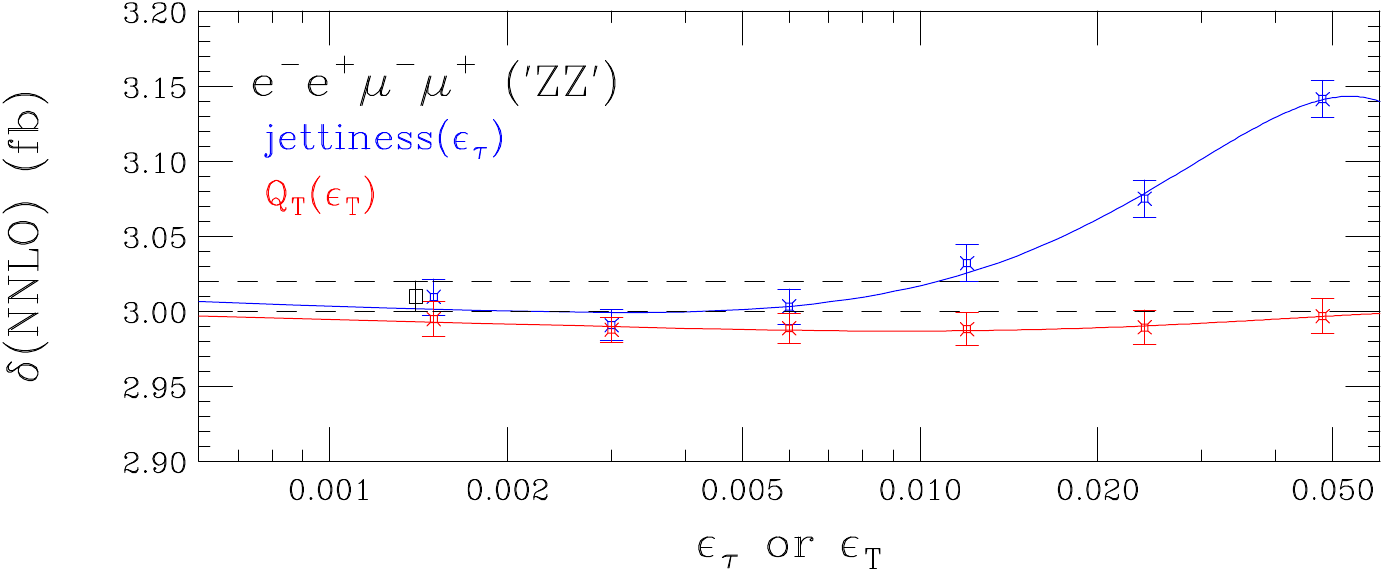}
\includegraphics[width=0.35\textwidth,angle=270]{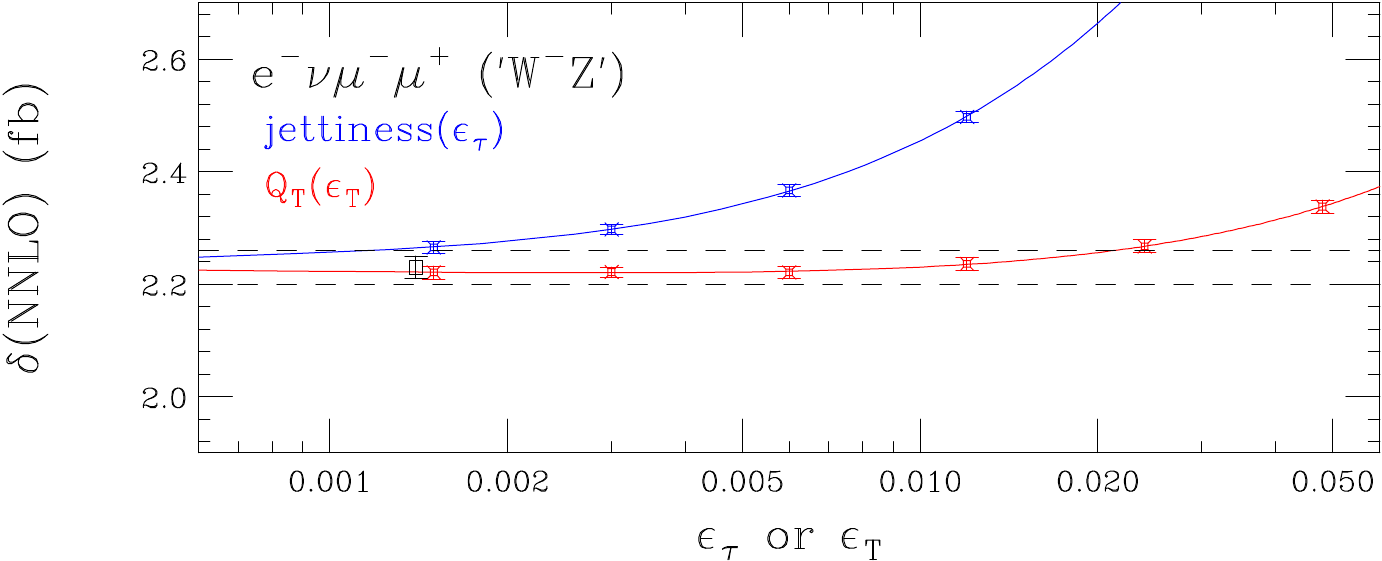}
\includegraphics[width=0.35\textwidth,angle=270]{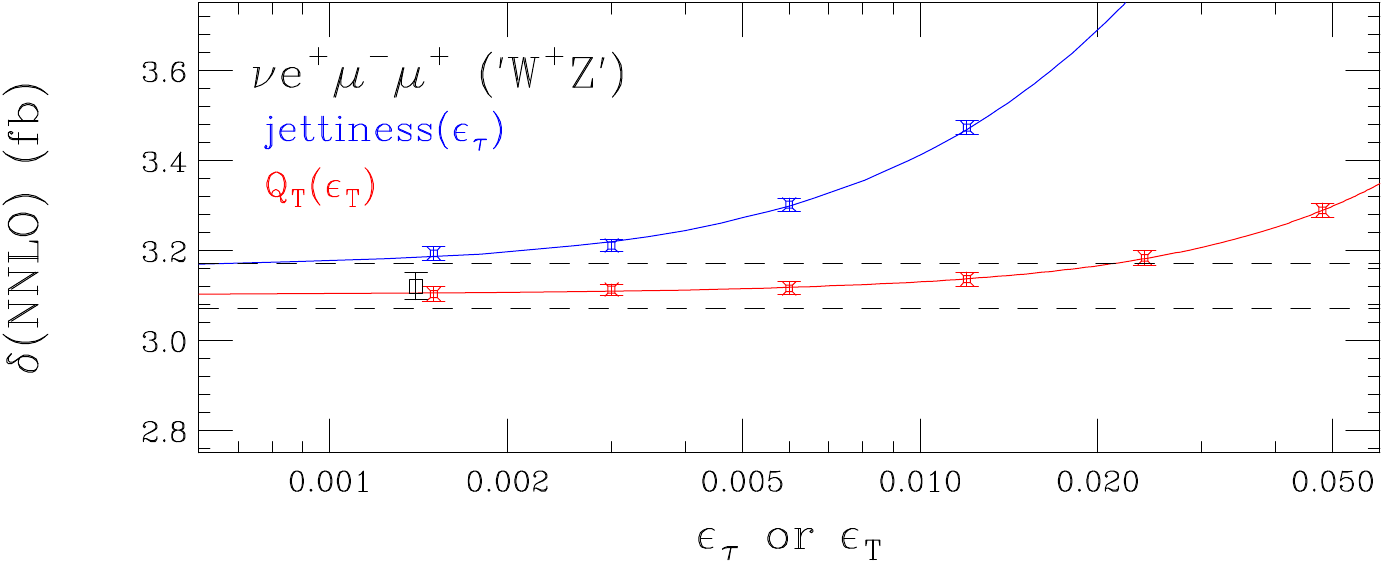}
\caption{Dependence of NNLO coefficient for $pp \to WW$, $pp \to ZZ$,
$pp \to W^-Z$ and $pp \to W^+Z$ processes on choice of slicing cut, for both
0-jettiness and $q_T$-slicing.
The MATRIX result for $q_T^{\rm cut} = 0.15\%$, Ref.~\cite{Grazzini:2017mhc} corresponds to the square black point (slightly offset
for visibility) and the uncertainty band of the extrapolated MATRIX result is shown as the dashed lines. 
\label{fig:dibosoncutdep}}
\end{figure}

Results for the NNLO corrections to the diboson processes considered in this
paper are shown in Figs.~\ref{fig:photoncutdep} and~\ref{fig:dibosoncutdep},
together with the benchmark results from table~\ref{tab:targets}
(extracted from  Ref.~\cite{Grazzini:2017mhc}). For the newly-included
processes in MCFM, shown in Fig.~\ref{fig:dibosoncutdep},  we note the
excellent agreement with the previous calculations reported by the MATRIX
collaboration.

As at NLO, for the processes with an identified photon in the final state the
approach to the asymptotic limit is similar for both $q_T$ and 0-jettiness
subtraction.  This indicates that power corrections to the factorization
theorems underlying Eqs.~(\ref{eq:SCETfac1}) and~(\ref{eq:dsigma_DY_full})
are affected by the requirement
of photon isolation in a similar way. For the other diboson processes $0$-jettiness suffers
from much larger power corrections than $q_T$ subtraction.

\subsection{Comparison with Ref.~\protect\cite{Heinrich:2017bvg}}
In Ref.~\cite{Heinrich:2017bvg} Heinrich et al have produced NNLO predictions
for $Z$-boson pair production using the 0-jettiness subtraction method
to isolate the doubly unresolved region. 
Note that in Ref.~\cite{Heinrich:2017bvg} the $Z$'s are considered on-shell,
and consequently there is no need to introduce the complex mass scheme. Adjusting
our input parameters accordingly,  we obtain the  results
shown in Table~\ref{tab:zzonshell}.  Excellent agreement with the earlier
calculation is observed.

\begin{table}[h]
\begin{center}
\begin{tabular}{ l|l|l|l}
& $\sigma_{LO}$ [pb] & $\sigma_{NLO}$ [pb] & $\sigma_{NNLO}$ [pb] \\ \hline 
Ref.~\cite{Heinrich:2017bvg} & $9.845$ & $14.100$ & $16.69(0)^{+3.1\%}_{-2.8\%}$ \\
MCFM                         & $9.856$ & $14.114$ & $16.68(1)^{+3.2\%}_{-2.7\%}$ \\
\end{tabular}
\caption{Comparison with the on-shell $ZZ$ results using NNPDF3.0 
from  Ref.~\cite{Heinrich:2017bvg}.  The quoted uncertainties at NNLO correspond
to those obtained by scale variation according to the procedure described in this
reference. \label{tab:zzonshell}}.
\end{center}
\end{table}

\clearpage

\section{Conclusion}
Our intent in the current paper has been to increase the range of
processes which are available in the MCFM package at NNLO in
QCD. Specifically we have added the processes, $W^+W^-$, $W^\pm Z$ and
$ZZ$, with leptonic decays of the $W$ and $Z$ bosons included.
As well as its extensive range of processes available at NLO
using dipole subtraction, MCFM now includes a wide range of processes
at NNLO. Representative results for the processes that are now included at
this order have been presented in Table~\ref{tab:targets}.
We have implemented two different slicing methods
for the calculation of the NNLO QCD corrections to processes with
colour singlet final states.  Both methods use global variables to
isolate the region of phase space with soft and collinear emission.

The jettiness method divides the phase space on the basis of the zero-jettiness,
defined in Eq.~(\ref{tau0CM}) whereas the $q_T$ method divides the phase space on
the basis of the total transverse momentum of the colour singlet particles.
We find that the $q_T$-slicing method appears to be subject to smaller power corrections
in most cases, although in certain cases ($W^\pm\gamma$, $Z\gamma$, $\gamma\gamma$ at NNLO)
the size of the power corrections is similar,
or even slightly smaller for jettiness. We note that all these processes involve photon isolation cuts.

Since there is a well-developed
literature on the summation of logarithms of transverse momentum the $q_T$-slicing method 
is easily extended to perform resummation of logarithms of $q_T$. For the moment the
application of the $q_T$-slicing method to coloured final states at NNLO has been
limited to the case of heavy-quark production~\cite{Catani:2019hip,Catani:2019iny,Catani:2020kkl}.
The extension to processes in which a massless parton is present in the final state is not straightforward,
although there are encouraging
signs at NLO using a $q_T$ surrogate, based on the $k_T$ jet algorithm~\cite{Buonocore:2022mle}. 
Despite the larger power corrections, which tend to disfavour the jettiness slicing method,
we note that the jettiness subtraction method already
has the proven ability to deal with coloured final states,
such as $W+{\rm jet}$, $Z+{\rm jet}$ and $H+{\rm jet}$.
Results for these processes have already been presented with MCFM.

We have provided a detailed
comparison of the two non-local subtraction methods in a way that they are directly comparable.
In this paper we have avoided as much as possible the imposition of cuts,
because of the influence that injudicious choices of cuts can have on the power corrections.
Resummation, by reducing the sensitivity to the low $q_T$ region,
has the benefit that it circumvents the additional power corrections which can occur
in the presence of some fiducial cuts~\cite{Alekhin:2021xcu,Salam:2021tbm}.
The calculations we have presented provide the basis for future extensions of MCFM that
include $q_T$ resummation for diboson processes, extending the work in
Refs.~\cite{Becher:2020ugp,Neumann:2021zkb}.
  
\paragraph{Acknowledgments}

We would like to thank Giuseppe de Laurentis for assistance simplifying one-loop matrix elements.
RKE thanks Gavin Salam for useful discussions.
This manuscript has been authored by Fermi Research Alliance, LLC under Contract No.
DE-AC02-07CH11359 with the U.S. Department of Energy, Office of Science, Office of High Energy
Physics.  The numerical calculations reported in this paper were performed using the
Wilson High-Performance Computing Facility at Fermilab.

\appendix
\section{Leading log behaviour of colour singlet production cross section}
\label{leadinglog}
For simplicity, we shall consider the simple Drell-Yan process in leading order,
although the discussion will apply {\it mutatis mutandis} to all colour singlet
final states. We follow closely the discussion of ref.~\cite{Ellis:1980my}.
The lowest order cross section has the form
\begin{equation}
\sigma(\shat,Q^2) = \sigma_0 \delta(1-z),\;\;z=Q^2/\shat,\;\;\sigma_0= \frac{4 \alpha \pi^2}{N \shat}.
\end{equation}
The invariant Drell-Yan cross section with the emission of one gluon of momentum $k$ is,
\begin{eqnarray}
\frac{\pi k^0 d\sigma}{d^3 k} &=&\sigma_0 
\int \; dx_1 dx_2 [f(x_1) f(x_2) + (1 \leftrightarrow 2)] \nonumber \\
&\times&   \frac{\alpha_s C_F}{2 \pi} 
    \Big(\frac{(\shat+\that)^2+(\shat+\uhat)^2}{\that \uhat}\Big) \delta( \shat+\that+\uhat-Q^2)\, ,
\end{eqnarray}
where $\shat=2 p_1.p_2$, $\that=-2p_1.k$, $\uhat=-2p_2.k$ and $k$ is the gluon momentum.
Taking the gluon momentum to be $k = \alpha p_1 +\beta p_2+ \vec{k}_T$
the invariant gluon momentum integral becomes,
\begin{equation}
  \frac{d^3 k}{k^0}= d \alpha \,d \beta \, d^2 \vec{k}_T \delta (\alpha \beta - \frac{\vec{k}^2_T}{\shat}).
\end{equation}
The parton cross section for fixed virtual photon transverse momentum $q_T$ is,
\begin{eqnarray}
  \frac{1}{\hat{\sigma}_0} \frac{d \sigma }{dq_T^2}
 &=& \,d \alpha \,d \beta \, 
 \delta (\alpha \beta - \frac{\vec{Q}^2_T}{\shat}) \delta(\shat(1-\alpha -\beta)-Q^2) \nonumber \\
  &\times &  \frac{\alpha_s C_F }{2 \pi} \Big[\frac{(1-\alpha)^2+(1-\beta)^2}{\alpha \beta}\Big].
\end{eqnarray}
To eliminate the delta functions in these expressions it is useful to
take moments with respect to $\rho =Q^2/\shat$ of the partonic cross section,
\begin{equation}
F_n(q_T^2/s)=\frac{1}{\hat{\sigma}_0} \int_0^1 d \rho \rho^{n+1} \frac{d \hat\sigma}{dq_T^2} .
\end{equation}
In the limit $q_T \to 0$ (ignoring powers of $\alpha,\beta$ in the numerator),
\begin{equation}
  F_n(q^2_T/\shat)=\frac{\alpha_s C_F }{\pi} \frac{1}{q_T^2} \int_{q_T^2/\shat} \frac{d\alpha}{\alpha}
   \sim -\frac{\alpha_s C_F }{\pi} \frac{\shat}{q_T^2} \ln\left(\frac{q_T^2}{\shat}\right).
\end{equation}
To explicitly exhibit the double logarithms we define,
\beq
\Sigma(q_T^{cut}/\shat)=\int_0^{{q_T^{cut}}^2} \, dq_T^2 F_n(q^2_T/\shat) .
\eeq
Integrating over $q_T$ up to $q_T^{cut}$ and cancelling IR singularities by inclusion of the virtual diagrams gives,
\begin{equation}
  \Sigma_T = \sigma_0 \Big[1 -\frac{\alpha_s C_F}{2 \pi} \ln^2 ((q_T^{cut})^2/Q^2)\Big]= \sigma_0
  \Big[1-\frac{2\alpha_s C_F}{\pi} \ln^2 (q_T^{cut}/Q)\Big]\,,
\label{eq:qtcut}
\end{equation}
the order $\alpha_s$ expansion of Eq.~(\ref{eq:SigmaT}).

This should be compared with the jettiness calculation,
\begin{eqnarray}
  F_n(\tau/\sqrt{\shat})&=&  \frac{\alpha_s C_F}{\pi} \int \frac{d \alpha}{\alpha}\int \frac{d \beta}{\beta}  \Big[ \theta(\alpha-\beta) \delta(\beta-\tau/\sqrt{\shat})+\theta(\beta-\alpha) \delta(\alpha-\tau/\sqrt{\shat})\Big] \\
  &=&  \frac{\alpha_s C_F}{\pi} \frac{\sqrt{\shat}}{\tau} \Big[ \int_{\tau/\sqrt{\shat}}^1 \frac{d \alpha}{\alpha} + \int_{\tau/\sqrt{\shat}}^1 \frac {d \beta}{\beta}\Big]\\
  &=&  -\frac{2 \alpha_s C_F}{\pi} \frac{\sqrt{\shat}}{\tau} \ln \frac{\tau}{\sqrt{\shat}} \, .
\end{eqnarray}
Integrating over $\tau$ up to $\tau^{cut}$ and cancelling IR singularities at $\tau=0$ by inclusion of the virtual diagrams gives,
\begin{equation}
\Sigma_\tau = \sigma_0 \Big[1-\frac{\alpha_s C_F}{\pi} \ln^2 \frac{\tau^{cut}}{Q}\Big]\, ,
\label{eq:taucut}
\end{equation}
the order $\alpha_s$ expansion of Eq.~(\ref{eq:Sigmatau}).

 \section{Translation of two-loop corrections to the hard function} 
\label{hardappendix}
For the $W^\pm\gamma, Z\gamma $ and $\gamma\gamma$ processes presented in Refs.~\cite{Gehrmann:2011ab,Anastasiou:2002zn}
the finite remainders of the two-loop matrix elements remove
singular terms of the form specified by Catani in Ref.~\cite{Catani:1998bh} but without a factor
of $(-\mu^2/s)^{2\epsilon}$ in the hard radiation factor ${\mathcal{H}}^{(2)}(\epsilon)$.
The translation from this scheme to a standard $\overline{\text{MS}}$ subtraction of the singularities, to obtain
the hard functions $H_{ij}$ introduced in section~\ref{sec:methods}, has been
described in Ref.~\cite{Becher:2013vva}.  The implementation of this conversion has been discussed in some detail
for these processes in Refs.~\cite{Campbell:2016yrh,Campbell:2017aul,Campbell:2021mlr}.\footnote{
arXiv:1603.02663v3 corrects typographical errors in previous versions of Ref.~\cite{Campbell:2016yrh}.}

For the diboson ($W^+W^-,W^\pm Z,ZZ$) processes presented in Ref.~\cite{Gehrmann:2015ora}
the finite remainders are presented in two schemes, in which the singularities are subtracted according either
to exactly Catani's scheme~\cite{Catani:1998bh} or to a scheme that is well-suited for the original
formulation of $q_T$ subtraction~\cite{Catani:2013tia}.
Starting from the latter ($\Omega^{(n),{\rm finite}}_{q_T}$), we convert to amplitudes that enter the hard
function ($\Omega^{(n),{\rm finite}}_{H}$) using the relations,
\begin{eqnarray}
\Omega^{(0),{\rm finite}}_{\rm H} &=& \Omega^{(0),{\rm finite}}_{q_T} \, , \nonumber \\
\Omega^{(1),{\rm finite}}_{\rm H} &=& \Omega^{(1),{\rm finite}}_{q_T} + \Delta I_1\, \Omega^{(0),{\rm finite}}_{q_T} \, , \nonumber \\
\Omega^{(2),{\rm finite}}_{\rm H} &=& \Omega^{(2),{\rm finite}}_{q_T} + \Delta I_1\, \Omega^{(1),{\rm finite}}_{q_T} + \Delta I_2\, \Omega^{(0),{\rm finite}}_{q_T} \, ,
\end{eqnarray}
where the coefficients are given by,
\begin{eqnarray}
\Delta I_1=            C_F   \left[            \frac{\pi^2}{12}          - \left(\frac{3}{2}          + i \pi \right) L          - \frac{L^2}{2}           \right] \,,
\end{eqnarray}
\begin{eqnarray}
\Delta I_2 &=& C_F^2 \left[\frac{\pi^4}{288} 
 + \left( - \frac{3}{8} + \frac{3\pi^2}{8} - 6 \zeta_3 - \frac{i \pi^3}{12} \right) L
 \right. \nn \\ && \left. \quad
 + \left(\frac{9}{8} - \frac{13 \pi^2}{24} + \frac{3 i \pi}{2} \right) L^2
 + \left(\frac{3}{4} + \frac{i \pi}{2} \right) L^3 + \frac{L^4}{8} \right] \nn \\ &&
 + C_F C_A \left( - \frac{607}{162} + \frac{67 \pi^2}{144} - \frac{\pi^4}{72} + \frac{77 \zeta_3}{36} + \frac{11 i \pi^3}{72} \right. \nn \\ && \left. \quad
 - \left(\frac{961}{216} + \frac{11 \pi^2}{36} - \frac{13 \zeta_3}{2} + \frac{67 i \pi}{18} - \frac{i \pi^3}{6} \right) L
 - \left(\frac{233}{72} + \frac{11 i \pi}{12} - \frac{\pi^2}{12} \right) L^2 - \frac{11 L^3}{36} \right] \nn \\ &&
 + C_F n_f (2 T_R) \left[\frac{41}{81} - \frac{5 \pi^2}{72} - \frac{7 \zeta_3}{18} - \frac{i \pi^3}{36} 
   + \left(\frac{65}{108} + \frac{\pi^2}{18} + \frac{5 i \pi}{9} \right) L
\right. \nn \\ && \left. \quad
   + \left(\frac{19}{36} + \frac{i \pi}{6} \right) L^2 + \frac{L^3}{18} \right] \,.
\end{eqnarray}
In these formulae the logarithm is $L=\log(\mu^2/s_{12})$ and analytic continuation has already been performed
assuming $s_{12} > 0$. As usual $C_F=4/3$, $C_A=3$ and $\zeta_3 = 1.20205690\ldots$.
Setting $L=0$ reproduces the conversion factors presented in Eq.~(2.9)
of Ref.~\cite{Heinrich:2017bvg}. After conversion in this way the amplitudes are suitable for implementation
in MCFM.

 \section{Cuts}
\label{appendixcuts}
\renewcommand\arraystretch{1.2}
\begin{table}
\begin{center}

\begin{tabular}{r|l}
{ photon cuts} 
& $\ptgammaone > 40$~GeV, $\ptgammatwo > 25$~GeV, $\etagamma<2.5$ \\
{ photon isolation} 
& Frixione isolation with $n=1$, $\varepsilon = 0.5$ and $\delta_0=0.4$ \\[-0.5cm] 
\\
{ jet definition}
& anti-$k_T$ algorithm with $R=0.4$;\; $\ptjet>25$~GeV, $|\eta_j|<4.5$
\end{tabular}
\end{center}
\renewcommand{\baselinestretch}{1.0}
\caption{\label{tab:gammagamma} Fiducial cuts for the $\gamma\gamma$ process.}
\vspace{0.75cm}
\end{table}

 \renewcommand\arraystretch{1.5}
\begin{table}
\begin{center}
\begin{tabular}{r | c | c}
& $pp\to e^-e^+\gamma$
& $pp\to e^-\bar\nu_e\gamma/pp\to e^+\nu_e\gamma$\\
\hline
\multirow{ 2}{*}{ lepton cuts} 
& $\ptlep > 25$~GeV, $\etalep<2.47$
& \multirow{ 2}{*}{$\ptlep > 25$~GeV, $\etalep<2.47$} \\[-0.1cm]
& $\mll > 40$~GeV
& \\
{ photon cuts}
& $\ptgamma > 15$~GeV, $\etagamma<2.37$
& $\ptgamma > 15$~GeV, $\etagamma<2.37$ \\
{ neutrino cuts}
&  n/a 
& $\ptmiss >35$~GeV\\
 separation cuts: &
\multicolumn{2}{c}{$\dRlepjet>0.3$, $\dRgammajet >0.3$,
$\dRlepgamma>0.7$}
\\
{ photon isolation}
& \multicolumn{2}{c}{Frixione isolation with $n=1$, $\varepsilon = 0.5$ and $\delta_0=0.4$}\\ 
{ jet definition}
& \multicolumn{2}{c}{anti-$k_T$ algorithm with $R=0.4$;\; $\ptjet>30$~GeV, $|\eta_j|<4.4$}
\end{tabular}
\end{center}
\renewcommand{\baselinestretch}{1.0}
\caption{\label{tab:Vgamma} Fiducial cuts for the
$pp\to e^-e^+\gamma$ ($Z\gamma$) and $pp\to e^-\bar\nu_e\gamma/pp\to e^+\nu_e\gamma$
($W^\pm\gamma$) processes.}
\vspace{0.75cm}
\end{table}
 \renewcommand\arraystretch{1.5}
\begin{table}
\begin{center}
\begin{tabular}{r | c | c}
& $pp \to e^- \mu^+ \nu_\mu \bar\nu_e $ 
& $pp\to e{\nu}_{e} \mu^+\mu^-$ \\
\hline
\multirow{ 3}{*}{ lepton cuts} 
& $\ptlone >25$~GeV, $\ptltwo >20$~GeV 
& $p_{T,\mu}>15$~GeV, $p_{T,e}>20$~GeV \\[-0.1cm]
& $|\eta_e|<2.47$, $|\eta_e|\notin[1.37;1.52]$
& $|\etal|<2.5$ \\[-0.1cm]
& $|\eta_\mu|<2.4$, $\mll>10$~GeV 
& $|m_{\mu^+\mu^-}-m_Z|<10$~GeV \\
{ neutrino cuts}
& $\ptmiss >30$~GeV, $\ptmissrel >15$~GeV 
& $\mtw> 30$~GeV \\
{ separation cuts}
& $\dRleplep>0.1$ 
& $\Delta R_{\mu\mu} >0.2$, $\Delta R_{\mu e}>0.3$ \\
{ jet cuts}
& $N_{\mathrm{jets}}=0$ & none \\[-0.5cm]
\\
{ jet definition} & \multicolumn{2}{c}{anti-$k_T$ algorithm with $R=0.4$;\; $\ptjet>25$~GeV, $|\eta_j|<4.5$}
\end{tabular}
\end{center}
\renewcommand{\baselinestretch}{1.0}
\caption{\label{tab:WWandWZ} Fiducial cuts for the
$pp \to e^- \mu^+ \nu_\mu \bar\nu_e $ ($WW$) and
$pp\to e{\nu}_{e} \mu^+\mu^-$ ($W^\pm Z$) processes.}
\vspace{0.75cm}
\end{table}
 \renewcommand\arraystretch{1.5}
\begin{table}
\begin{center}
\begin{tabular}{r | l }
{ lepton cuts}
& $\ptlep > 7$~GeV, $\etalep<2.7$, $66{\rm~GeV}<\mll<116$~GeV \\
{ separation cuts}
& $\dRleplep>0.2$ \\
{ jet definition}
& anti-$k_T$ algorithm with $R=0.4$;\; $\ptjet>25$~GeV, $|\eta_j|<4.5$
\end{tabular}
\end{center}
\renewcommand{\baselinestretch}{1.0}
\caption{\label{tab:ZZ} Fiducial cuts for the 
$pp \to e^- e^+ \mu^- \mu^+$ ($ZZ$) process.}
\vspace{0.75cm}
\end{table}
 The fiducial cuts for the $\gamma\gamma$ production process
are given in Table~\ref{tab:gammagamma},
for $Z\gamma$ and $W^\pm\gamma$ production processes in Table~\ref{tab:Vgamma},
for $WW$ and $W^\pm Z$ production processes in Table~\ref{tab:WWandWZ},
and for $ZZ$ processes in Table~\ref{tab:ZZ}.
These cuts have been deliberately chosen to be the same as Ref.~\cite{Grazzini:2017mhc}.

 \bibliography{main}
\bibliographystyle{JHEP}
\end{document}